\documentclass[namedreferences]{solarphysics}
\usepackage[optionalrh]{spr-sola-addons} 
\usepackage{epsfig}          
\usepackage{graphicx}        
\usepackage{natbib}         
\usepackage{amssymb}        
\usepackage{color}           
\usepackage[T1]{fontenc}        

\newcommand{\caxix}{{Ca~{\sc xix}}}
\newcommand{\caxviii}{{Ca~{\sc xviii}}}

\newcommand{\fexxi}{{Fe~{\sc xxi}}}
\newcommand{\fexxii}{{Fe~{\sc xxii}}}
\newcommand{\fexxiii}{{Fe~{\sc xxiii}}}
\newcommand{\fexxiv}{{Fe~{\sc xxiv}}}
\newcommand{\fexxv}{{Fe~{\sc xxv}}}
\newcommand{\fexxvi}{{Fe~{\sc xxvi}}}

\begin{document}

\begin{article}

\begin{opening}

\title{New Results from the {\em Solar Maximum Mission Bent Crystal Spectrometer}}

\author{C.G.~\surname{Rapley}$^{1}$\sep
        J.~\surname{Sylwester}$^{2}$\sep
        K.J.H.~\surname{Phillips}$^{3}$\sep
             }
\runningauthor{C.G. Rapley {\it et al.}}
\runningtitle{New Results from the SMM/BCS}

\institute{$^{1}$ Dept. of Earth Sciences, University College London, Gower Street, London WC1E 6BT, UK email: \url{christopher.rapley@ucl.ac.uk} \\
$^{2}$ Space Research Centre, Polish Academy of Sciences, 51-622, Kopernika~11, Wroc{\l}aw, Poland
                     email: \url{js@cbk.pan.wroc.pl} \\
$^{3}$ Scientific Associate, Earth Sciences Department, Natural History Museum, London SW7 5BD, UK\\
                     email: \url{kennethjhphillips@yahoo.com}\\}
\date{\today}


\begin{abstract}
The {\em Bent Crystal Spectrometer} (BCS) onboard the NASA {\em Solar Maximum Mission} was part of the {\em X-ray Polychromator}, which observed numerous flares and bright active regions from February to November 1980, when operation was suspended as a result of the failure of the spacecraft fine pointing system. Observations resumed following the  Space Shuttle SMM Repair Mission in April 1984 and continued until November 1989. BCS spectra have been widely used in the past to obtain temperatures, emission measures, and turbulent and bulk flows during flares, as well as element abundances. Instrumental details including calibration factors not previously published are given here, and the in-orbit performance of the BCS is evaluated. Some significant changes during the mission are described, and recommendations for future instrumentation are made. Using improved estimates for the instrument parameters and operational limits, it is now possible to obtain de-convolved, calibrated spectra that show finer detail than before, providing the means to improved interpretation of the physics of the emitting plasmas. The results indicate how historical, archived data can be re-used to obtain enhanced and new, scientifically valuable results.
\end{abstract}
\end{opening}

\section{Introduction}\label{Intro}

The diagnostic power and scientific value of high-resolution X-ray spectra emitted by hot (1~MK to $\approx 30$~MK) astrophysical plasmas have been well established over the past half-century. Several solar spacecraft-borne X-ray spectrometers were flown in the 1960s and 1970s which showed the presence of prominent spectral lines due to the H-like and He-like ions of abundant elements in the solar coronal active regions and flares \citep{neu70,dos72,wal71,gri73,par72}.

Until 1980, spacecraft-borne X-ray spectrometers used the conventional principle in which plane crystals diffract incoming X-rays according to the Bragg condition (see Equation~(\ref{Bragg}) below).  Rotating the crystals such that the angle of incidence varies with time generates a complete spectrum. For solar flares, the time scales of variability can be very short relative to the spectrometer scan duration, which has generally been significantly greater than ten seconds. As a result, the source variability is folded into the scanned spectrum, a considerable disadvantage during the flare's impulsive stage for which the variability time scales are only a few seconds. This can be overcome with crystals that are slightly curved, such that the entire spectral range can be dispersed and instantaneously registered with a position-sensitive detector. This ``bent crystal'' technology was introduced for a prototype instrument successfully launched on an Aerobee rocket \citep{cat74,rap77}. Subsequently, an instrument with eight such curved crystals was proposed for inclusion on {\em Solar Maximum Mission}\/ (SMM) and accepted as part of the {\em X-ray Polychromator} (XRP) package. An early (and pre-launch) outline of the instrument was given by \cite{act80}. The instrument produced numerous results over the SMM lifetime (1980\,--\,1989) apart from a period when the spacecraft attitude-control unit, ensuring fine Sun pointing, was inoperable (November 1980 to April 1984). Analyses based on spectra from the SMM/BCS have already been published, with a much improved understanding of flare plasmas as a consequence \citep{bel82b,lem84,ant82}. At present, standard analysis programs for the BCS written in the Interactive Data Language (IDL) are available through the Solar Software (SSW) package \citep{fre98}.

For the purposes of this article, we have derived improved BCS crystal and instrument parameters using both pre-launch and in-orbit measurements. The revised parameters have been used to de-convolve the spectra, and as a result flare spectra with greatly improved resolution have now been obtained.  Many details of the spectra, including bulk motions of plasma at the flare impulsive stage and the satellite line structure, are now more evident. In this article, examples of such improved spectra are given, as well as instrumental details including calibrations not previously published. We assess the instrument's in-orbit performance including limits beyond which data become unreliable, some instrument calibration drifts and degradation, and changes revealed in images obtained by the Space Shuttle astronauts during the 1984 Repair Mission.

\section{The BCS Instrument} \label{Instrument}

\subsection{Principles}

The principle of bent crystal spectrometers is based on the Bragg diffraction equation,

\begin{equation}
n \lambda = 2d \,{\sin} \, \theta {\rm ,}
\label{Bragg}
\end{equation}

\noindent where $n$ is the diffraction order, $\lambda$ the wavelength, $d$ the crystal lattice spacing, and $\theta$ the glancing angle of incidence. The BCS viewed only first-order ($n=1$) spectra. For a crystal curved along its length, incident X-rays intercept the crystal at angles that differ slightly depending on their arrival position. The entire dispersed spectrum can be registered with a position-sensitive detector, with the temporal resolution set by the encoding time of the detector and its electronics. A collimator is necessary to define the instrument boresight and constrain its field of view, since the flares that occur off-axis in the dispersion plane are displaced along the detector axis. Similarly, spectra from sources extended within the field of view are broadened as a result of the ambiguity between incidence-angle and wavelength.

In addition to its scientific advantage of providing high temporal resolution, the BCS design offers reduced electromechanical complexity since no scan drive or angle-encoding mechanisms are required. However, these advantages are achieved at the cost of increased electronic complexity, and a significantly higher data rate. In addition, care must be taken to ensure satisfactory pre-flight adjustment of the collimators, crystals, and detectors, and adequate in-flight stability of the instrument geometry. Small deviations of the geometry from the ideal case, such as variations of the crystal curvature, translate into distortions of the spectral dispersion. Similarly, spatial non-uniformities of the heat shield and collimator transmissions, crystal reflectivity, or detector sensitivity translate into differential variations in the instrument sensitivity along the dispersion axis. Depending on their magnitude, such anomalies can distort the spectra recorded, potentially compromising the plasma-diagnostic estimates derived. Applying corrections is not necessarily straightforward, since as noted above spectra from flares that occur off-axis in the plane of dispersion are shifted relative to the pattern of anomalies.

\subsection{Instrument Details and Pre-Launch Measurements}

\subsubsection{BCS Design and Elements}

\begin{figure}
\centerline{\includegraphics[width=0.5\textwidth,clip=,angle=90]{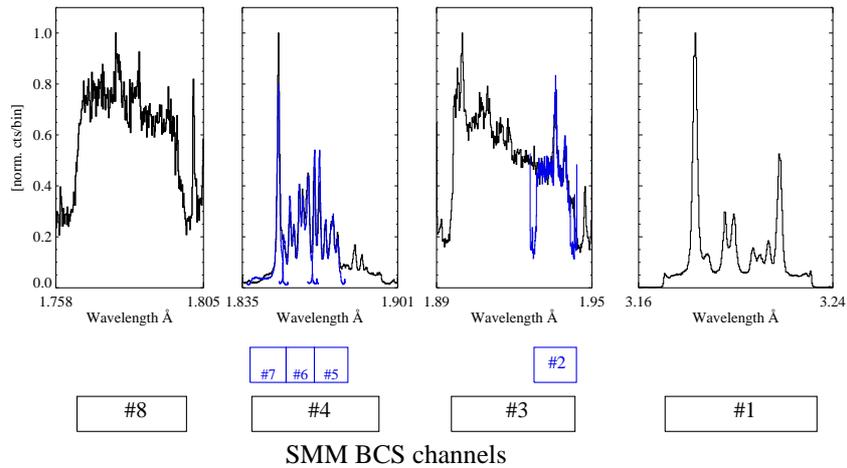}}
\caption{Wavelength coverage of the eight BCS channels shown as blocks with BCS spectra from the peak of the 01~July 1980 (SOL1980-07-01T16:28) flare shown to illustrate each channel's wavelength range. Spectra from channels 1, 3, 4, and 8 are shown in black, and those from channels 2, 5, 6, and 7 are in blue.}\label{spectra_wavelength_ranges}
\end{figure}

In accordance with the SMM spacecraft objectives, among which was the study of the development of flares and other dynamic phenomena, the BCS was designed to have eight channels covering groups of spectral lines emitted by high-temperature plasmas. Seven of these (channels 2\,--\,8) covered line emission in the range 1.78\,--\,1.94~\AA\ due to iron, in particular the lines of H-like Fe (\fexxvi, 1.78~\AA) and He-like Fe (\fexxv) and associated dielectronic satellite lines (1.85\,--\,1.90~\AA) as well as the K$\alpha$ doublet lines (1.92\,--\,1.94~\AA) produced by the fluorescence of neutral iron in the photosphere. These choices were driven by the then recent results \citep{neu70,dos71,gri73} showing the considerable diagnostic potential of intensity ratios of these lines. BCS channel~1 covered the group of lines due to He-like Ca (\caxix) and associated satellites at $\approx 3.2$~\AA. Table~\ref{BCS_channels} gives nominal wavelength ranges and spectral lines of the eight BCS channels together with crystal parameters including the range of Bragg angles at the extreme ends of each crystal. Figure~\ref{spectra_wavelength_ranges} shows the wavelength ranges of each channel with reference to BCS spectra for a GOES X2.5 flare on 01~July 1980 (peak 16:28~UT: SOL1980-07-01T16:28) seen by SMM instruments.

\begin{table}
\caption{Design parameters of SMM/BCS channels}
\label{BCS_channels}
\begin{tabular}{llccccc}
\hline                   
Channel & Ion line  & Nominal   & Bragg angles & Wavelength  \\
No.\tabnote{Crystals: Ge 220 for channel~1 ($2d = 4.000$ \AA); Ge 422 for channels 2\,--\,8 ($2d=2.310$ \AA)}    & emission\tabnote{Indicates range of ions with strong lines emitted during flares. LO = low resolution; HI = high resolution.}   & Curvature & [deg.]    & Range [\AA]\tabnote{For an on-axis source.}  \\
    && radius [m] \\
  \hline
1 & \caxix\ $w$-$z$ lines & 5.83 &  52.29\,--\,53.86 & 3.165\,--\,3.231 \\
2 & Fe K$\alpha$ HI  & 10.26 &  56.59\,--\,57.36 & 1.928\,--\,1.945 \\
3 & {\fexxi}\,--\,Fe K$\alpha$ LO & 3.56 &  55.04\,--\,57.42 & 1.893\,--\,1.947 \\
4 & {\fexxv}\,--\,\fexxii\ LO & 3.99 & 52.81\,--\,55.08 & 1.840\,--\,1.894  \\
5 & {\fexxiv} sats. HI & 16.52  & 53.89\,--\,54.44 & 1.866\,--\,1.879  \\
6 & {\fexxiv} sats. HI & 14.85 & 53.39\,--\,53.93 & 1.854\,--\,1.867  \\
7 & \fexxv\ $w$ line HI & 15.13& 52.89\,--\,53.43 & 1.842\,--\,1.855  \\
8 & \fexxvi\ Ly-$\alpha$ lines & 7.30 & 49.99\,--\,51.04 & 1.769\,--\,1.796  \\
  \hline
\end{tabular}
\end{table}

The BCS consisted of the following components:

\noindent i) Heat shield;

\noindent ii) Multi-grid Oda collimator;

\noindent iii) Eight dispersive diffracting X-ray crystals;

\noindent iv) Eight position-sensitive proportional counters with commandable $^{55}$Fe calibration sources (two sources per deck of four detectors);

\noindent v) Analogue electronics and high-voltage units;

\noindent vi) Digital electronics.

\begin{figure}
\centerline{\hspace*{0.015\textwidth}
               \includegraphics[width=0.43\textwidth,clip=,angle=0]{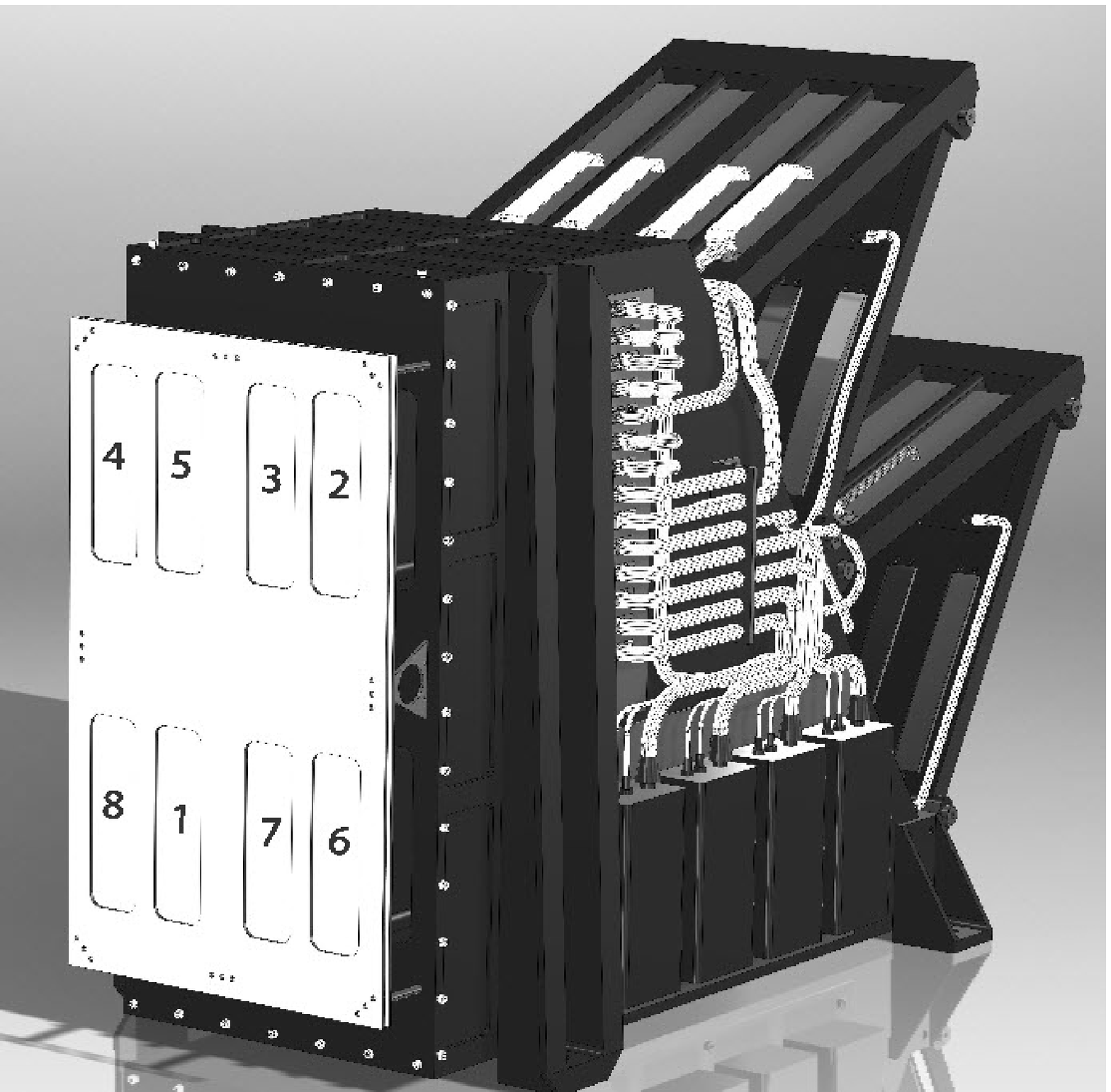}
               \hspace*{0.03\textwidth}
               \includegraphics[width=0.5\textwidth,clip=]{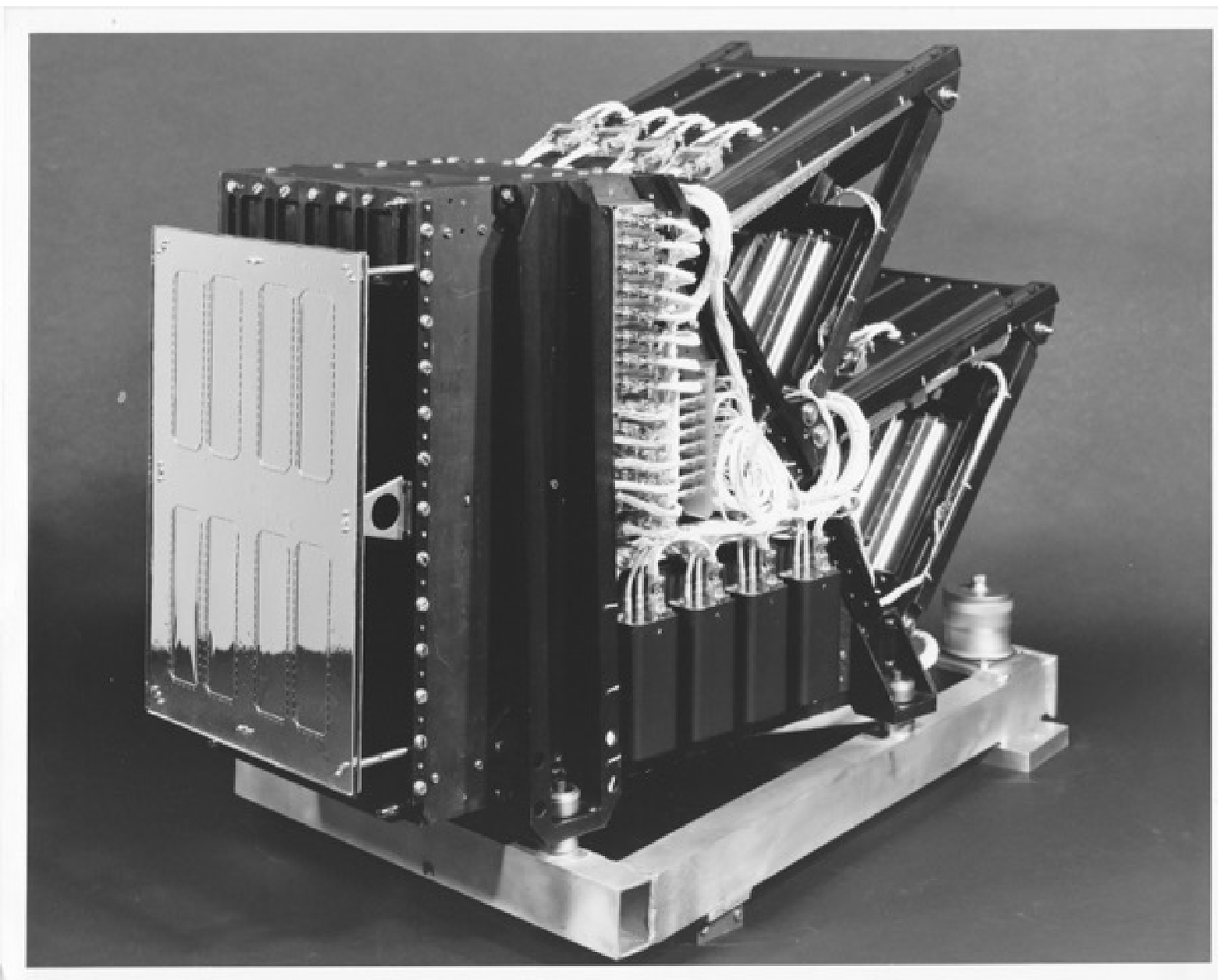}}
\caption{Left: CAD illustration of the SMM/Bent Crystal Spectrometer with channel numbers indicated. Right: The instrument bolted to a calibration jig prior to installation in the SMM spacecraft. The heat shield can be seen mounted on the instrument front face (to the left). The analogue-signal-processing electronics, digital electronics, and power supplies were mounted elsewhere in the spacecraft and are not shown. The instrument height was 580~mm and width 350~mm. }\label{BCS_before_launch}
\end{figure}

Figure~\ref{BCS_before_launch} gives two views of the instrument, with the computer-aided design (CAD) illustration in the left panel showing the crystals (in colour) and the position-sensitive detectors. The eight channels, which are numbered in the figure, were arranged in two decks of four arranged one above the other. The right panel of Figure~\ref{BCS_before_launch} shows the BCS before spacecraft launch, with the heat shield covering the collimator apertures. The analogue-signal-processing electronics, digital electronics, and electrical power supplies were mounted separately within the spacecraft. Apart from the radioactive calibration sources, the instrument contained no moving parts. The structure was of machined magnesium to provide mechanical rigidity and thermal stability, but with low mass (total instrument mass was 35~kg). The structure was fixed via a three-point kinematic mount to the spacecraft instrument support plate, with nominal alignment of better than one~arcmin to the {\em Fine Pointing Sun Sensor} in pitch and yaw. As the normal orientation of spacecraft was zero roll, positive yaw corresponded to solar East direction and positive pitch corresponded to solar South direction. In terms of the standard solar coordinate system used by {\em Solar and Heliospheric Observatory}, {\em Solar Dynamics Observatory}, and other spacecraft \citep{tho06}, these directions correspond to $-x$ and $-y$ respectively. From here on, we will use the standard ($x$, $y$) coordinates instead of yaw and pitch, which were more familiar during the SMM operations period.

\subsubsection{Heat Shield}
\label{Heat_Shield}

The heat shield was designed to prevent thermal distortions of the collimator. It was a key component of the overall thermal design, the objective of which  was to achieve a uniform and stable instrument temperature close to 20$^\circ$C, so that pre-flight calibrations and set-up remained valid. It consisted of two elements: an outer Second Surface Mirror (SSM) constructed from a 25.4$\mu$m-thick Teflon film with a 1000~\AA\ aluminium coating on the inner surface (away from the Sun), and a 12.7$\mu$m-thick Kapton thermal membrane (TM) which also had a 1000~\AA\ aluminium coating on its inner surface. The films were mounted on rigid aluminium frames.

The heat-shield transmission was measured pre-launch using an $^{55}$Fe source and laboratory X-ray detector. The results were consistent with the thickness data supplied by the manufacturers, which had been verified by mechanical spot checks.

Photo-absorption data for the heat shield material \citep{Hen93} indicate that at a wavelength of $\approx 3.2$~\AA\ (that of the \caxix\ lines viewed in channel~1: photon energy 3.87~keV) the combined transmission of the SSM and Kapton inner film was 47\,\%. This value is lower than the 65\,\% quoted previously. The lower value is adopted here. For the remaining BCS channels viewing lines from Fe ions, the combined transmission lay in the range 83\,\% to 88\,\%.

\subsubsection{Collimator}
\label{collimator}

The collimator was constructed in two halves, one mounted above the other, and each illuminating a deck of four spectrometers. Each half comprised two identical 200-mm long magnesium optical benches mounted within the instrument structure. On each optical bench were seven magnesium frames arranged in a geometric progression based on the standard Oda principle (Oda, 1965). Each frame contained four $120 \times 25$~mm apertures, which supported 83\,$\mu$m-thick electroformed nickel grids. The grid holes were squares with sides 350\,$\mu$m and with 215\,$\mu$m bars. This gave a field of view on the Sun nominally $6 \times 6$ arcminutes (FWHM) square, the intention being to exclude emission from flares from neighbouring active regions that would have caused spectral confusion. The peak transmission was in principle 38.4\,\%, assuming a perfect alignment of the nine serial grids. In practice, the grid alignment tolerance was $\pm 10\,\mu$m; assuming an aggregate misalignment of 20\,$\mu$m, the equivalent hole size was 330\,$\mu$m, giving a theoretical peak transmission of 34.1\,\%. The grid bars were 20\,\% larger than for a perfect Oda collimator to ensure that any distortions or X-ray transparency of the rounded edges would not result in significant side-lobe transmission.

Pre-launch measurements using a copper K$\alpha$ line source (8.1~keV) in a 17.5-m-long vacuum tube indicated that the collimator field of view was $6.0 \pm 0.6$ (FWHM)~arcmin with a mean peak transmission of $33 \pm 3$~($1\sigma$)\,\%, sampled at three points for each of the eight apertures. Variations along the dispersion axis were less than or equal to 10\%. Both collimator assemblies were mounted on the instrument structure with measured co-alignment better than $\pm 30$~arcsec; the co-alignment with the SMM spacecraft boresight reference mirror was also better than $\pm 30$~arcsec. Measurements indicated that side-lobe transmission was insignificant: only 0.5\,\% out to $\pm 54$~arcmin off-axis.

\subsubsection{Crystals}
\label{sec:crystals}

The choice of diffracting crystals was determined by the wavelengths covered, reflectivity, and spectral resolution. Resolving powers [$\lambda / \Delta \lambda$] of approximately $3000 - 6000$ were required for the satellite-line structure to be visible. Germanium offered an acceptable compromise between high reflectivity and spectral resolution, with Ge~220 for channel~1 (\caxix, $2d = 4.000$~\AA), and Ge~422 ($2d = 2.310$~\AA) for the remaining channels. The crystals were supplied by Quartz et Silice (France) and were cut from a single boule of high-quality germanium. The crystal dimensions were 160~mm $\times$ 30~mm and thickness 1~mm. The exact length of crystal illuminated depended on the geometry of each spectrometer and ranged from 143~mm (channel 2) to 155~mm (channel 8). Each crystal was supplied with the large faces ground flat and parallel to within 5~arcmin. Following further grinding and chemical etching, individual crystals were screened using an X-ray technique to select those with the highest spectral resolution. Each was then mounted against a precision mandrel fitted with a space-approved elastomer gasket and a metal clamping frame to assume the required curvatures with radii in the range 3.5~m to 16.5~m. In order to reduce distortions introduced by inaccuracies in the figure of the mandrel only the perimeter of each crystal was mechanically supported. In practice, several cycles of X-ray measurements and re-lapping of the mandrels were necessary to achieve sufficiently uniform degrees of curvature. The gaskets avoided damage to the crystals during the launch vibration, and increased the tolerances on the clamping frames. Each crystal was mounted at an angle to the instrument boresight corresponding to the mean Bragg angle within the range covered (see Table~\ref{BCS_channels}).  Pre-launch alignments were achieved using detachable optical reference mirrors.

\begin{figure}
\centerline{\includegraphics[width=0.95\textwidth,clip=,angle=0]{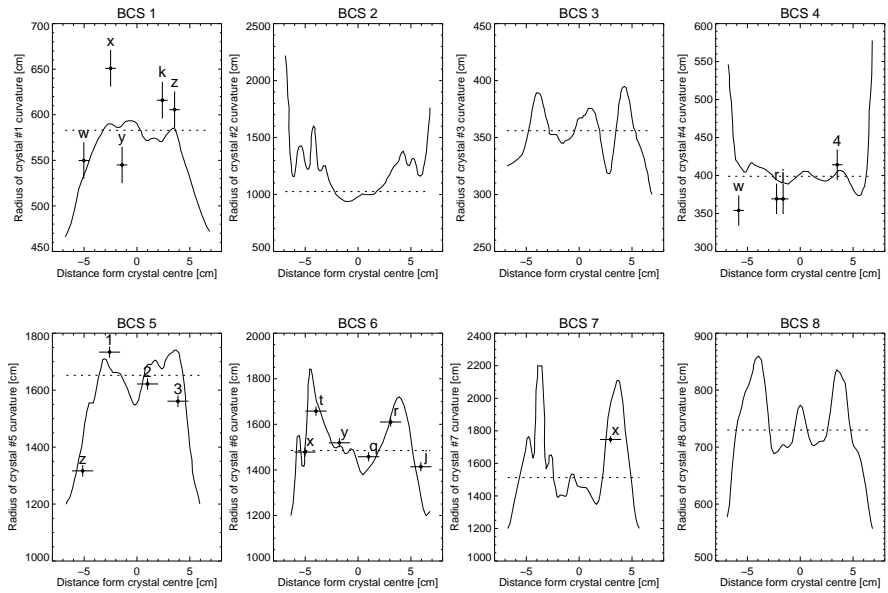}}
\caption{Radius-of-curvature measurements (solid curves) made during the qualification of the eight BCS crystals (channel numbers are indicated above each plot). The crosses indicate radius measurements made using spectral lines in each channel (notation of \cite{gab72}) during the spacecraft raster of 06~November 1980. The nominal (design) curvature for each crystal is shown as a dotted horizontal line in each panel. Figure adapted from \cite{parm81}.}\label{crystal_bend_radii_APThesis}
\end{figure}

The crystal bend radii were determined by finding the centre of crystal rocking curves every 5~mm along four tracks across the 25-mm crystal width. Data for the central two tracks were summed to give crystal radius as a function of position along the crystal. For five of the crystals, these showed variations of up to $\pm 10$\,\% over the portions of the crystal illuminated by solar X-rays and larger variations towards the crystal ends. However, the crystals for channels~2, 6, and 7 showed substantially larger variations over parts of their surfaces. Figure~\ref{crystal_bend_radii_APThesis} (solid curves) shows the measured radii as a function of position along the crystal. Nominal radii, given in Table~\ref{BCS_channels}, are also shown (dotted lines) in the figure.

The variations in crystal curvature along the dispersion axis are significant since the spectral dispersion is inversely proportional to the crystal radius. The effect is ``elastically'' to stretch or compress the solar spectrum, distorting the spectral lines so that the count rates in a given digital bin are unchanged. However, the integral count rate under the line profile is increased or decreased proportionately. (See Section~\ref{sec:wave_calib} for in-orbit measurements of crystal radii in the vicinity of strong spectral lines, and estimates of the average radii over the full spectral range.)

\subsubsection{Detectors}

The one-dimensional position-sensitive proportional counter detectors were of a type described by \cite{bor68}. The lids and bodies were machined under numerical control from medium-strength aluminium alloy with thicknesses as low as 1~mm to reduce mass. The internal cross sections were 20~mm (depth) by 26~mm (width), and the window apertures were 134~mm long, of which 121\,--\,125~mm were illuminated by solar X-rays. The small differences were a consequence of the differing divergence of the beams diffracted from each crystal. The divergence was greatest for channel~3 with the largest crystal curvature, and least for channel~5. The windows were of beryllium foil with thickness 140\,$\mu$m for the Fe channels (2\,--\,7) and 90\,$\mu$m for the Ca channel (channel~1). The foils extended over the full length of the detector and were sealed with an indium cord set in a precisely machined ``O'' ring groove. The windows were supported over the 25~mm $\times$ 134~mm aperture with a 10-mm deep aluminium alloy honeycomb of 3~mm hexagonal cells, limiting radiation with incident angle to within 18$^\circ$ of the normal. The on-axis transmission of the honeycomb was 90\,\%. A 2-mm slot was cut into the support lid on each side of the aperture (157~mm apart) to admit 5.9~keV radiation from the $^{55}$Fe X-ray calibration sources. The anodes were made of 50\,$\mu$m diameter carbon-coated quartz (manufactured by Carl Zvanut Co., USA) with a resistance of 1.5\,M$\Omega$~m$^{-1}$.

The detectors were filled to 900 torr pressure with a 49\,\% Ar, 49\,\% Xe gas mixture with 2\,\% carbon dioxide as the quenching gas. The detectors operated at 2500~V with a gas gain of approximately $10^4$. The position resolution of each detector at 5.9~keV (described by a Gaussian with a small degree of skewness towards the detector ends) was 0.9~mm (FWHM). The skewness was minimised by mounting each detector tilted such that X-rays in the centre of the spectral range were received at normal incidence.

\subsubsection{Analogue Electronics}
\label{anal_elec}

The analogue electronics were housed at the back of the detectors, with signals transmitted via co-axial cables (11~m in length) to the separately located digital position-encoding electronics and microcomputer, where the data were processed, formatted, and passed to the spacecraft telemetry system.
The detector resistive anode and end capacitances corresponded to an RC integrator modifying the leading edges of the pulses detected at each end of the anode in a manner that depended on the X-ray photon arrival position: the sum of the pulses corresponded to the output of a conventional proportional counter. The position encoding is described by \cite{rap77}. Opposite ends of each detector anode were connected to a pair of low-noise, high-impedance, voltage-sensitive pre-amplifiers, the output of which were fed to amplifiers and pulse-shaping filters, and then to the pulse summing and position-encoding electronics. The difference in crossover times for the analogue pulses from each end of the anode corresponding to a single photon were linearly related to the distance (to within 1\,\%) between the detection distance of the photon event and the centre of the detector anode for suitably chosen pulse-shaping double differentiation and single integration time constants (2\,$\mu$s for the BCS). A single-channel pulse height analyzer (PHA) applied to the summed pulses from both detector ends rejected cosmic ray background events outside its range. Detector gains were monitored in-orbit regularly with an eight-channel PHA analyzing signals from the $^{55}$Fe calibration sources. With the sources exposed, the position data from the calibration slots allowed the position registration, scale, and resolution to be monitored. An onboard test pulse generator was provided to check the functionality and stability of the electronics, and to mark the ``electronic'' central digital bin of each detector system.

\subsubsection{Digital Electronics}
\label{dig_elec}

The photon-arrival positions in channels 1, 2, 3, 4, and 8 were digitized into a sign bit and a seven-bit word, equivalent to 256 digital bins, with a 25\,$\mu$s encoding time. For the higher spectral resolution channels 5, 6, and 7 the photon-arrival positions were digitized into 128 digital bins with a 13\,$\mu$s encoding time. Fast emitter coupled logic (ECL) circuitry was used for the sign-bit determination to ensure correct encoding of events close to the electrical centre of the anode (see, however, Subsection~\ref{det_iss}). In order to reduce power consumption, all other circuitry consisted of discrete low-power CMOS components. At the end of the conversion period for an event, the data were presented via a parallel interface to the BCS microcomputer system with a ``valid data present'' signal. The system remained locked out to further pulses until read by the microcomputer. An inter-event timer was incorporated to prevent pulse pile-up from generating erroneous data when the interval between pulses was shorter than 40$\mu$s.  This limited the operating range of the detector to a count rate of order 5000 s$^{-1}$, beyond which counts would be progressively lost and position encoding increasingly distorted.

A pioneering innovation in the BCS was the inclusion of a microprocessor (an RCA 1802 with 12K words of memory) to control the instrument and to process its data (a hard-wired backup controller was provided in case of failure). This allowed full advantage to be taken of the unique ability of the instrument to provide very high temporal resolution spectra. Spacecraft telemetry limitations (approximately 152 eight-bit words~s$^{-1}$ or 1218 bits s$^{-1}$) dictated that although the BCS microcomputer could generate spectra from all eight channels in 128\,ms, readout of the output of all eight BCS channels with full spectral coverage required a time of 7.65~seconds. Higher temporal resolution was possible through selection of the spectral coverage and resolution. Thus, by summing the counts in bins outside those covering the solar spectra to provide background information, a temporal resolution of 6.3~seconds could be achieved. Degrading the spectral resolution by summing position bins allowed a temporal resolution of as short as 128\,ms.

A feature of the BCS that enabled data to be accepted at very high rates was a 64K memory called the Queue memory. A total-counts scalar could be steered to any of the eight detectors, its output being made available to the microcomputer for processing to generate a flare flag from a selected BCS channel or the SMM/{\em Hard X-ray Imaging Spectrometer} (HXIS). This was used to control the data formatting mode in a variety of ways. In the simplest case, a flare trigger could be generated in response to a count rate exceeding a pre-selected threshold. Other, more sophisticated tests were available including double or multiple threshold checks to give a flare-rise-time monitoring capability. On detection of a flare, data could be generated at a rate in excess of the telemetry rate for period of time, ranging from seconds to many minutes, depending on the observing mode, using the Queue as a buffer. Multiple changes of temporal resolution or spectral format were possible, to match the temporal evolution of flares ({\em e.g.} slow precursor, rapid rise, slow decay). Alternatively, the Queue could be used in a ``circular'' mode, rapidly overwriting until the detection of a flare, at which time any part of the data within the queue could be ``frozen'' and read out before subsequent lower-rate data, thereby providing a high-time-resolution data stream immediately pre-flare.

In practice, the two most commonly used observing modes were ``All bins: 7.65~s time resolution'' and ``All bins: 3.8~s time resolution'', the latter being triggered by the flare flag and operating for seven minutes, by which time the Queue was full.

The programming of the BCS microprocessor was carried out in hexadecimal machine code and offered considerable flexibility. The ability to upload new code during the mission provided the means to adjust the instrument operating modes in the light of experience, and to respond to instrument anomalies. The {\em Solar Maximum Mission} was one of the first scientific spacecraft in which microprocessors were ubiquitous throughout the instrument complement and spacecraft.

\section{BCS Characteristics}

\subsection{Sensitivity: Effective Areas}

The BCS sensitivity is here expressed by the instrument's effective area [$A_{\rm eff}$: mm$^2$] which converts photon count rate [$C$: counts s$^{-1}$] summed over the range of bins corresponding to each spectrometer's resolution element to absolute fluxes [$F$: photons cm$^{-2}$ s$^{-1}$] by $F = C / A_{\rm eff}$. The bin range corresponds to the convolution of crystal rocking curve, detector position resolution, and digital resolution (see Section~\ref{sec:wave_res} and Table~\ref{BCS_wave_res}). The effective area for a BCS channel is the product of the crystal area for each channel exposed to solar X-rays [$A$\/] (which varies slightly from channel to channel due to the differing angles of incidence of the incoming collimated beam on the crystal), the crystal integrated reflectivity [$R_{\rm int}$: radians], factors $\kappa_{\rm geom}$  expressing the combined transmission of the collimator and detector honeycomb, $\kappa_{\rm det}(\lambda)$ the detector window transmission, $\kappa_{\rm SSM}(\lambda)$ the SSM transmission, and $\kappa_{\rm TM}(\lambda)$ the TM transmission (see Section~\ref{Heat_Shield}), all divided by the range of Bragg angles of each channel. Thus, the effective crystal area for each BCS channel (central wavelength $\lambda$) is defined to be

\begin{equation}
A_{\rm eff}(\lambda) = \frac{A \sin \theta}{\Delta \theta} \,\,\frac{180}{\pi}\,\, R_{\rm int}\,\,(\lambda)\,\,\kappa_{\rm geom} \,\,\kappa_{\rm det}(\lambda)\,\, \kappa_{\rm SSM}(\lambda) \kappa_{\rm TM}(\lambda)
\label{eff_xtal_area}
\end{equation}

\noindent where $\Delta \theta = \theta_1 - \theta_2$ is the range of Bragg angles [degrees] over the crystal. Over the small wavelength interval of each channel, all appropriate parameters may be considered constant, so to a first approximation $A_{\rm eff}(\lambda)$ is a single number for each channel. For integrated reflectivities, measured values for the SMM/{\em Flat Crystal Spectrometer} (FCS: Acton et al., 1980) Ge~220 and Ge~422 crystals were used (Section~\ref{sec:wave_res}). The transmission factor [$\kappa_{\rm geom}$] is equal to the product of the collimator transmission ($33 \pm 3$\,\% for boresight axis, all channels) and that of the detector honeycomb (90\,\%, all channels). Values of the transmission factors [$\kappa_{\rm det}(\lambda)$,  $\kappa_{\rm SSM}(\lambda)$, and $\kappa_{\rm TM}(\lambda)$] are given in Table~\ref{BCS_eff_areas}. Those for the SSM and TM include the transmission of the 1000~\AA\ aluminium coatings. The nominal effective areas [$A_{\rm eff}$] in Table~\ref{BCS_eff_areas} are refinements of earlier estimates given by \cite{act80} (Table~3), although the revisions are only slight except for channel~1 where the smaller value reflects the revised pre-launch estimate of heat shield transmission (Section~\ref{Heat_Shield}), and for channels~2 and 8 where it has not been possible to reconcile the previous values. Also shown in Table~\ref{BCS_eff_areas} are effective areas $A_{\rm eff-io}$ based on in-orbit estimates of $\Delta \theta$. These differ by up to $\approx 30$\,\% from the pre-launch values. Note that following the loss of the Second Surface Mirror in October 1980, the $A_{\rm eff}$ values should be increased by a factor $1/\kappa_{\rm SSM}$.

\begin{table}
\caption{BCS Wavelength Resolution}
\label{BCS_wave_res}
\begin{tabular}{llccccc}
\hline                   
Chan.   & Crystal       & Detector    & Digital   & Combined   & Combined     \\
No.     & Rocking       & resolution  & bin width & Resolution & Resolution    \\
        & Curve FWHM    & FWHM        & [m\AA]    & (FWHM)     & (FWHM)  \\
        & [m\AA]\tabnote{Values are for FCS flat crystals, assumed to be those for BCS crystals (see text).}&         [m\AA]  &         & [m\AA]     & [bins] \\
  \hline
1       & 0.628 & 0.422 & 0.303 & 0.764 & 2.518\tabnote{This value for 1980; for April 1985: 2.612 bin; for April 1987: 3.158 bin.}  \\
2       & 0.067 & 0.122 & 0.070 & 0.137  & 1.957  \\
3       & 0.068 & 0.389 & 0.232 & 0.405  & 1.744  \\
4       & 0.068 & 0.389 & 0.258 & 0.404  & 1.670  \\
5       & 0.068 & 0.096 & 0.127 & 0.121  & 0.949  \\
6       & 0.069 & 0.096 & 0.133 & 0.120  & 0.902  \\
7       & 0.070 & 0.096 & 0.142 & 0.121  & 0.852  \\
8       & 0.076 & 0.194 & 0.183 & 0.215  & 1.173  \\
  \hline
\end{tabular}
\end{table}

\begin{table}
\resizebox{\textwidth}{!}{
\caption{BCS Effective Areas}
\label{BCS_eff_areas}
\begin{tabular}{llllllllllll}
\hline                   
Ch.   & Mid- & Bragg  & sin~$\theta$ & $R_{\rm int}$ & $\kappa_{\rm det}$ & $\kappa_{\rm TM}$ & $\kappa_{\rm SSM}$ & $L$ & $A$ & Eff. & $A_{\rm eff-io}$ \\
No.     & chan.  & angle &       & $10^{-5}$ rad    &     &      &  & mm & mm$^2$& area & mm$^2$ \\
        & wvl.   & range &&&&&&&& $A_{\rm eff}$   \\
        & [\AA]        &$\Delta \theta$ &&&&&&&& mm$^2$ \\
\hline
\multicolumn{8}{l}{For all channels $\kappa_{\rm geom} = 0.33 (\pm 0.03)\times 0.90$} \\
1       & 3.198   & 1.57 & 0.799 & 10.5 & .85 & .895 & .523 & 150.2 & 3755 & 1.357 & 1.574 \\
2       & 1.937   & 0.77 & 0.838 & 3.5  & .95 & .977 & .866 & 143.2 & 3580 & 1.954 & 2.599 \\
3       & 1.920   & 2.38 & 0.831 & 3.3  & .95 & .977 & .871 & 144.4 & 3610 & 0.594 & 0.766 \\
4       & 1.867   & 2.27 & 0.809 & 2.9  & .95 & .979 & .879 & 148.5 & 3712 & 0.539 & 0.625 \\
5       & 1.873   & 0.55 & 0.811 & 3.0  & .95 & .979 & .878 & 148.0 & 3700 & 2.305 & 2.628 \\
6       & 1.861   & 0.54 & 0.806 & 2.9  & .95 & .978 & .880 & 148.9 & 3722 & 2.258 & 2.484 \\
7       & 1.849   & 0.54 & 0.800 & 2.8  & .95 & .980 & .882 & 150.0 & 3750 & 2.174 & 2.217 \\
8       & 1.783   & 1.05 & 0.772 & 2.3  & .95 & .981 & .894 & 155.4 & 3885 & 0.899 & 0.728 \\
  \hline
\end{tabular}
}
\end{table}

For flares not on the BCS boresight, the collimator transmission [$T$] is progressively smaller than the measured on-axis value of 33\,\%. For channel~1, the fractional collimator transmission [$T$] for a flare offset only in the BCS dispersion plane such that the \caxix\ line~$w$ at bin number $B$ is

\begin{equation}
T = 0.33 \times \frac{ 6.0  - |B - 189.0| \times 0.46}{6.0}.
\label{trans_line_w}
\end{equation}

\noindent Collimator transmission corrections for flares offset orthogonal to the dispersion plane can in principle be estimated using data from the FCS, the {\em Ultraviolet Spectrometer/Polarimeter} (UVSP), or ground-based imagery. Data from the HXIS are unfortunately no longer accessible.

\subsection{Wavelength Response}
\label{sec:Wavelength_Response}

The wavelength response of each spectrometer was set by the design geometry, the precision and accuracy with which it was achieved during assembly, and any shifts or distortions that occurred through launch or in orbit. The pre-flight alignment of the crystals to the collimator boresight was achieved to $\pm 30$~arcsec using optical reference mirrors. The deviations from perfection of the crystal curvatures have been discussed in Section~\ref{sec:crystals} and will be addressed further in Section~\ref{sec:wave_cal}. The detector position responses were calibrated pre-flight and were monitored in orbit using the $^{55}$Fe calibration sources and position calibration slots.

\subsection{Wavelength Resolution}
\label{sec:wave_res}

Pre-launch crystal measurements showed rocking-curve widths (FWHM) of $\approx 54$~arcsec for the Ge~220 crystal (channel~1) and $\approx 11$~arcsec for the Ge~422 crystal (channels 2\,--\,8), the shapes being approximately Lorentzian. Rocking curves and integrated reflectivities were separately measured (by A.~Burek) pre-launch for the XRP/{\em Flat Crystal Spectrometer} (FCS) crystals including the Ge~220 and Ge~422 crystals (used for FCS channels~6 and 7 respectively); the rocking-curve widths for these crystals are only a few percent different from those of the BCS crystals. Confirmation that this was so was obtained from an on-line software toolkit for X-ray optics (\textsf{XOP}: \citep{xop04}) giving rocking curves and integrated reflectivities for Ge~220 and Ge~422 flat crystals and bent crystals with bend radii equal to those of the BCS crystals and both were found to be indistinguishable. The theoretical rocking-curves were calculated to be between 0.4 and 0.6 times those of the measured values, a typical fraction for high-quality (but still imperfect) crystals. Table~\ref{BCS_eff_areas} (column 5) gives the Burek measured integrated reflectivities [$10^{-5}$ radians] and Table~\ref{BCS_wave_res} (column 2) the rocking curve widths [FWHM in m\AA].

The position resolution of each detector [$\Delta L$\/] was measured in the laboratory at 10-mm intervals along its length using an $^{55}$Fe (5.9 keV) source and narrow mechanical collimated slot. The distribution was Gaussian with FWHM equal to 0.9~mm at the detector centre, but with slight (outward) skewness towards the detector ends. Tests showed a small reduction in FWHM for lower energy X-rays, estimated to be 10\,\% at 3 keV. With the exposed length of the detector anode [$L$] corresponding to 121\,--\,125~mm, the detector resolution in wavelength measure is approximately $(\Delta L / L) \times \Delta \lambda$ where $\Delta \lambda$ is the wavelength range of each channel, and $\Delta L$ is 0.8~mm for channel~1 and 0.9~mm for all other channels.  Values are listed in Table~\ref{BCS_wave_res} (column 3). Table~\ref{BCS_wave_res} (column 4) also gives the digital bin widths [m\AA] determined in-orbit in 1980 using the $^{55}$Fe sources and position calibration slots to derive the conversion from bins to mm, and in-orbit fits to observed spectra to obtain the conversion from bins [mm] to m\AA. The profile of each digital bin can be regarded as a rectangle with width as listed.

The (Gaussian) detector resolution and the (near-Lorentzian) rocking-curve profile were combined to form a Voigt profile, and this profile was combined with the rectangular-bin-width profile to give a final value for the spectral resolution (FWHM) for each channel in bins and m\AA. These values are given in the final two columns of Table~\ref{BCS_wave_res}. Numerical values in units of bins of less than 1 (channels~5, 6, and 7) are a mathematical artefact of the overall fit to the broad wings of the Voigt profile.


\section{In-Orbit Performance}
\label{sec:inorbitperformance}

\subsection{History}

The SMM launch took place successfully from Cape Kennedy, Florida, USA on 14~February 1980. The BCS turn-on and check-out were completed between the third and fifth days in orbit. The first (weak class C) flare was observed in channel~1 (\caxix) on 25~February. For the next nine months, except for brief interruptions unassociated with the BCS, the instrument provided continuous coverage of the SMM-selected active region for the 60 sunlit minutes of each 94-minute orbit. Operating modes were coordinated with the other instruments onboard the spacecraft, with those on other spacecraft (notably the US Naval Research Laboratory's {\em Solar Flare X-rays} (SOLFLEX) and {\em Solar X-ray} (SOLEX) instruments on the US Air Force {\em P78-1} spacecraft: \cite{dos79,mck81}), and with ground-based observatories by an international team of scientists at the Experiment Operations Facility at NASA's Goddard Space Flight Center. The spacecraft attitude control system failed in November 1980, as a result of which precise pointing at active regions became impossible and the BCS, like the other pointed instruments onboard SMM, was turned off.

Observations recommenced in April 1984, following the repair of the SMM by astronauts on the NASA Shuttle Rescue Mission STS-41-C.  Almost immediately some of the most intense flares observed during the mission occurred, despite the solar cycle being four years into its declining phase. Over the nine-year mission, many hundreds of flares of GOES class C or above were recorded. The last observation was made on 24~November 1989, after which the spacecraft re-entered the Earth's atmosphere, burning up on 2~December.

The BCS design life was a minimum of one year with a maximum of seven years. Its performance was generally excellent, exceeding its design criteria. However, a few failures occurred, and some significant instrument artifacts were revealed, which are discussed in the following sections.

\subsection{In-flight Calibrations}
\label{sec:wave_calib}

\subsubsection{BCS Detector Linearity}
\label{sec:det_linear}

Pre-launch measurements of the BCS detector linearities (bin number {\em versus} distance along the anode)
carried out using an $^{55}$Fe source collimated with a 1~mm slot were consistent with those from laboratory prototypes and the detector flown previously on an Aerobee payload \citep{rap76,rap77}. Departures from linearity were characterized by a sine wave of wavelength equal to the anode length and amplitude less than 1\,\%.
In-orbit position calibrations using the on-board $^{55}$Fe sources radiating through the calibration slots over the nine-year duration of the mission revealed a small but significant increase in some detector position gains, and a changing offset of the central digital bin. This accounted for a small spreading and shifting of the corresponding spectra recorded. The effect was greatest in detectors~1 and 7 and least in detectors~4 and 6.

\begin{figure}
\centerline{\includegraphics[width=1.0\textwidth,clip=,angle=0]{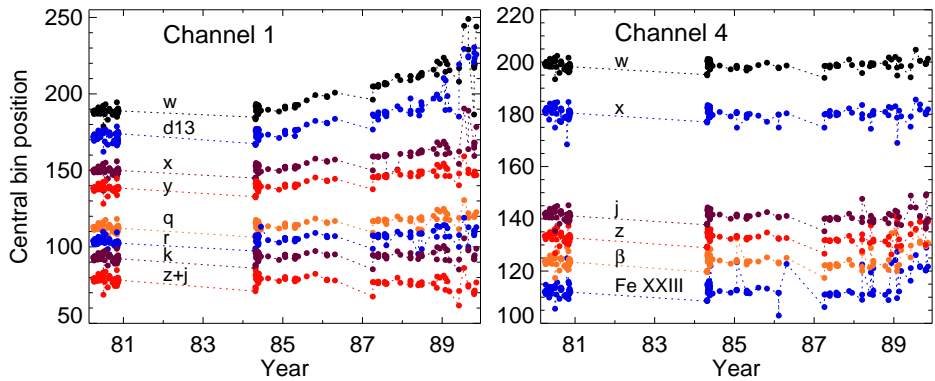}}
\caption{Left panel: Fitted bin positions of \caxix\ line $w$ and other lines in BCS channel~1 {\em versus} time over the lifetime of SMM. Right panel: Corresponding data for \fexxv\ line $w$ and other lines in channel~4. (Line notation follows \cite{gab72}.)}
\label{fitted_ch1_ch4_lines_vs_time}
\end{figure}

Analysis of 144 of the strongest flares through the mission shows the evolution of the digital bin values for prominent lines in channels~1 and 4 (Figure~\ref{fitted_ch1_ch4_lines_vs_time}). The changes in position gain and spectral shift in channel~1 are evident. The data indicate that the change corresponds to a linear ``stretch''. This was almost certainly caused by a change in the product of RC (detector anode resistance and end capacitance). Several possible explanations exist for such a change, but are unresolvable without laboratory tests which are no longer feasible. Life testing of the BCS detectors before launch was limited by the mission-preparation schedule and by the strengths of laboratory radioactive sources available, so no changes of this type were detected. For future missions, the use of more intense sources and extended life tests are recommended.

Meanwhile, linear fits applied to the BCS flight spectra compensate the effect (see Section~\ref{sec:wave_cal}). The fits are made to spectra from the early decay stage of flares, giving an optimum bin\,--\,wavelength relation for each flare. Flares were chosen such that there were no detectable Doppler shifts of transverse motions and prominent lines with well-established wavelengths were visible (see discussion in Section~\ref{sec:wave_cal}).

\subsubsection{Wavelength calibration}
\label{sec:wave_cal}

Given the closely linear outputs of the BCS detectors, and assuming uniformly curved crystals (this is refined later in this section), the wavelength calibration of each BCS channel can be taken as approximately linear,

\begin{equation}
\lambda = m B + \lambda_0
\label{wave_linear_cal}
\end{equation}

\noindent where $m$ is a constant, $B$ is the bin number, $\lambda$ is wavelength, and $\lambda_0$ is the wavelength corresponding to bin 0 for a flare on axis ({\em i.e.,} aligned with the collimator boresight in the dispersion plane).

For channels 1 to 7, determinations of the boresight were made in-orbit during an M3.5 flare on 06~November 1980 (peak time 22:27~UT: SOL1980-11-06T22:27). While the flare was in progress, the spacecraft executed a series of angular raster manoeuvres in the solar E\,--\,W ($\pm x$ ) direction, which was the BCS dispersion axis. These scans consisted of angular displacements of about $\pm 3.5$~arcminutes with a cycle time of 240~seconds, during which time the BCS accumulated spectra with 6.3-second time resolution. As a result of the scanning motion, the spectra appeared to ``slide'' back and forth along the dispersion axis as the angle of incidence of the incoming beam on the crystals varied.

\begin{figure}
\centerline{\includegraphics[width=1.0\textwidth,clip=,angle=0]{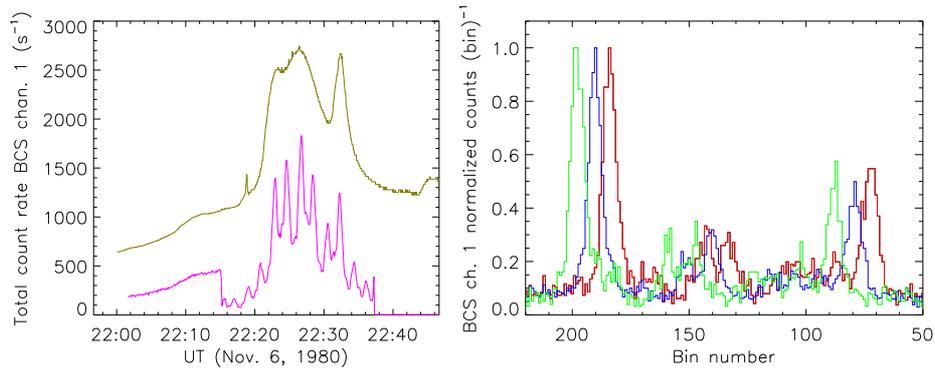}}
\caption{Left panel: BCS light curve (in magenta: total counts s$^{-1}$, channel~1) during the spacecraft scans of 06~November 1980 (22:15\,--\,22:37 UT) when an M3.5 flare (SOL1980-11-06T22:27) was in progress; modulations are due to the flare repeatedly passing through the BCS collimator response. Normalized GOES ($0.5-4$~\AA) light curve plotted in olive. Right panel: Normalized BCS channel~1 spectra plotted against bin number at 22:23:48~UT (red), 22:24:39~UT (blue), and 22:25:36~UT (green), illustrating the shift of spectra against bin number for when the flare was along the BCS boresight (blue) and at each end of the spacecraft manoeuvre range (red and green). The accumulation time for each spectrum was 6~seconds, and the angular range of the spacecraft scans was $\pm 3.5$~arcmin in the BCS dispersion direction.}
\label{Nov6_1980_lc_spch1}
\end{figure}

Figure~\ref{Nov6_1980_lc_spch1} (left panel) shows the GOES $0.5-4$~\AA\ light curve and the BCS channel~1 light curve; the modulation of the latter by the collimator response function during each scan is evident. Figure~\ref{Nov6_1980_lc_spch1} (right panel) shows channel~1 \caxix\ spectra for the extremes of the angular scans and on the boresight. The peak times of the BCS light curves are assumed to be those when the flare appeared along the BCS boresight, and hence spectra at these times are those for which values of $m$ and $\lambda_0$ can be determined. These are given in Table~\ref{wave_cal}. Prominent lines with known wavelengths (from \cite{kel87} or \cite{bea67} in the case of the Fe K$\alpha$ lines) are required for the calibration, as indicated in the table. The high-temperature \fexxvi\ Ly$\alpha$ lines (1.778, 1.784~\AA: \cite{kel87}) were not detectable in channel~8 during the 06~November flare. However, they were visible during the X3.3 flare of 14~October 1980 (peak time 06:11~UT: SOL1980-10-14T06:11), which data from the other BCS channels demonstrated occurred on the instrument boresight. The calibration constants for channel~8 were derived in this way.

\begin{table}
\caption{BCS Wavelength Calibration}
\label{wave_cal}
\begin{tabular}{lllcll}
\hline                   
Chan.   & Lines used        & Line & $m$\tabnote{Channels~1\,--\,7 from flare of 06~November 1980 (22:27:04 UT); channel~8 from 14~October 1980 (06:07:35~UT).}   & $\lambda_0$        & Boundaries [bins]    \\
No.     & for calibration   & wavelength [\AA]\tabnote{Line wavelengths from Kelly (1987) apart from Fe K$\alpha$ lines (Bearden 1967).}  & [m\AA] & [\AA]   \\
  \hline
1       & \caxix\ $w$       & 3.1769          & - 0.305  & 3.2343             & 28.0\,--\,223.5 \\
2       & Fe K$\alpha_1$, K$\alpha_2$ & 1.9360, 1.9400  & - 0.0094 & 1.948              & 34.5\,--\,221.5 \\
3       & Fe K$\alpha_1$, K$\alpha_2$ & 1.9360, 1.9400  & - 0.2521 & 1.951              & 29.5\,--\,228.5 \\
4       & \fexxv\ $w$       & 1.8503          & - 0.2580 & 1.901              & 26.0\,--\,223.0 \\
5       & \fexxv\ $w$       & 1.8503          & - 0.1264 & 1.879              & 15.5\,--\,113.5 \\
6       & \fexxv\ $x$, $y$  & 1.8553, 1.8594  & - 0.1437 & 1.869              & 19.5\,--\,115.5 \\
7       & \fexxv\ $z$       & 1.8680          & - 0.1420 & 1.856              & 16.5\,--\,112.5 \\
8       & \fexxvi\ Ly$\alpha_1$, $\alpha_2$  & 1.778, 1.784 & - 0.185 & 1.806 & 43.5\,--\,219.5 \\
  \hline
\end{tabular}
\end{table}

\begin{figure}
\centerline{\includegraphics[width=0.95\textwidth,clip=,angle=0]{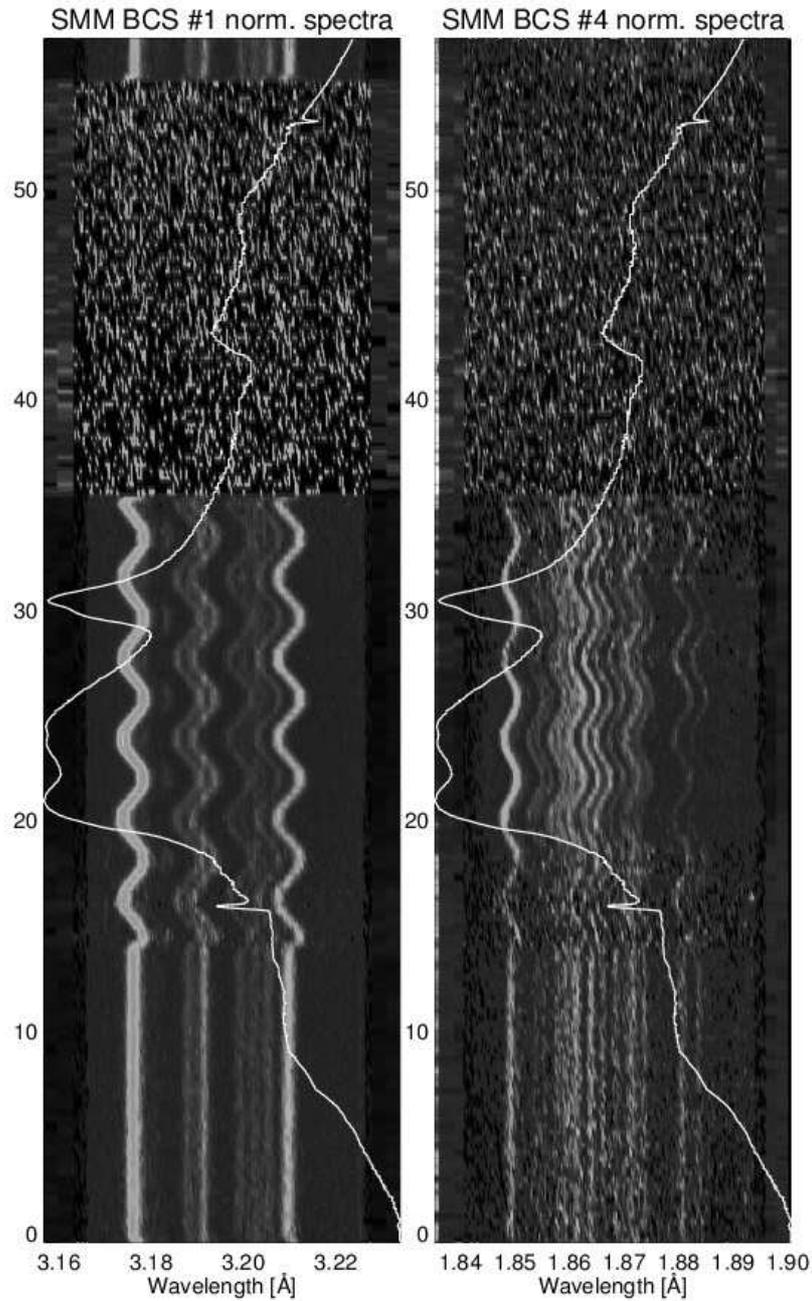}}
\caption{Spectral\,--\,time plots for BCS channels 1 (left) and 4 (right) for the 06~November 1980 M3.5 flare (see Figure~\ref{Nov6_1980_lc_spch1}) with intensities indicated by a red temperature colour scale. The wavelength scale is shown at the foot of each panel and the time scale (in minutes from 22:01:47~UT) along the vertical axis. The spacecraft scans appear as wavelength modulations in spectral lines within each channel. The white line is the  GOES 1\,--\,8~\AA\ flux plotted against time. }
\label{Nov6_1980_sp_time_scans}
\end{figure}

Figure~\ref{Nov6_1980_sp_time_scans} shows BCS spectral data from the 06~November 1980 flare on a colour intensity scale representation for channels~4 (\caxix) and 7 (\fexxv\ LO), with wavelength along the horizontal axis and time along the vertical axis. The wavelength modulations due to the spacecraft scans here appear as five cycles of sine waves in the spectral lines over the 22:15\,--\,22:37~UT period.

\begin{figure}
\centerline{\includegraphics[width=0.75\textwidth,clip=,angle=0]{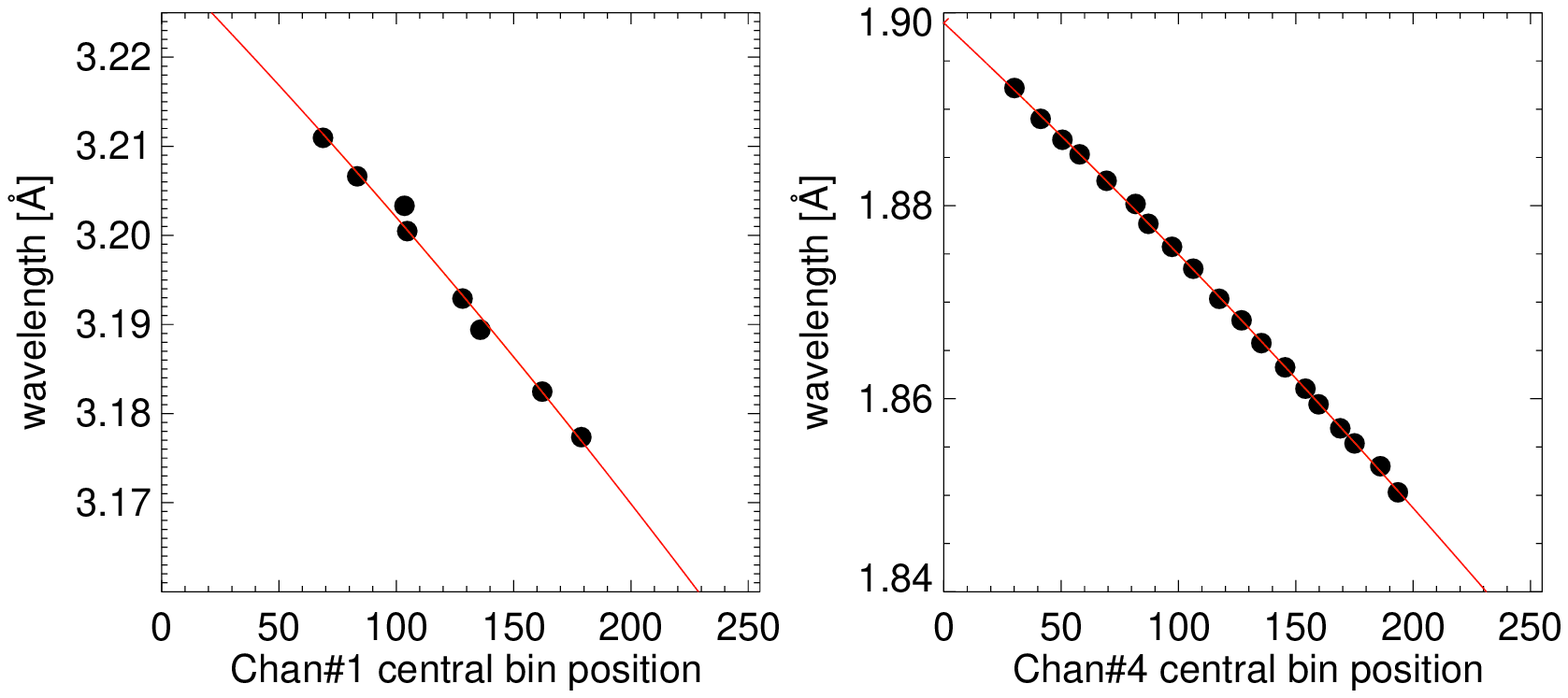}}
\caption{Bin number of the peak of prominent spectral lines {\em versus} line wavelength for BCS channels~1 (left) and 4 (right) during the 01~July 1980 flare (SOL1980-07-01T16:28) to illustrate the near-linearity of these channels during the 1980 Mission. The spectral lines include those shown in Figure~\ref{fitted_ch1_ch4_lines_vs_time}.}
\label{wvl_bin_plot}
\end{figure}

As noted in Section~\ref{sec:det_linear}, departures from spectrometer linearities (Equation~(\ref{wave_linear_cal})) were examined from in-orbit data by finding the relation between theoretical wavelengths for prominent spectral lines, generally from \cite{kel87}, and the detector bin locations in which they occurred. Figure~\ref{wvl_bin_plot} shows the results of an analysis of the GOES X2.5 flare on 01~July 1980 (SOL1980-07-01T16:28). The channel~1 plot is derived using the \caxix\ lines and \caxviii\ satellites, and the channel~4 plot uses the \fexxv\ lines and \fexxiv\ satellites. The relation is clearly close to linear but the departures from linearity ($< 5$\,\% in channel~1 and $< 2$\,\% in channel~4) are important for measurements of the wavelengths of lines falling in these two channels.

Pre-launch calibrations showed that the crystal curvatures deviated from constant values by up to $\pm 10$\,\% (larger in channels~2, 5, and 7) over the region illuminated by solar X-rays (Section~\ref{sec:crystals}) resulting in non-uniformities in spectral dispersion. \cite{parm81} took advantage of the 06~November 1980 spacecraft scans to investigate this, exploiting the fact that the migration of a spectral line in response to a given angular offset is directly related to the crystal curvature at the point on the crystal at which the diffraction occurs. By measuring the differing rates of migration for prominent lines as a function of angular offset, he was able to estimate the radii of curvature at the corresponding crystal locations. The results are plotted in Figure~\ref{crystal_bend_radii_APThesis}, in which the in-orbit estimates are shown as points with error bars relative to the pre-flight laboratory calibrations shown as continuous curves. The results for channels~5, 6, and 7 confirm the pre-flight calibrations to within the measurement uncertainties. The results for channels 1 and 4, with higher crystal curvatures, show some deviations, but they lie within the 20\,\% range measured pre-flight. It was not possible to carry out an analysis for channels~2, 3, and 8 as the spectra were too weak.

\begin{figure}
\centerline{\includegraphics[width=0.95\textwidth,clip=,angle=0]{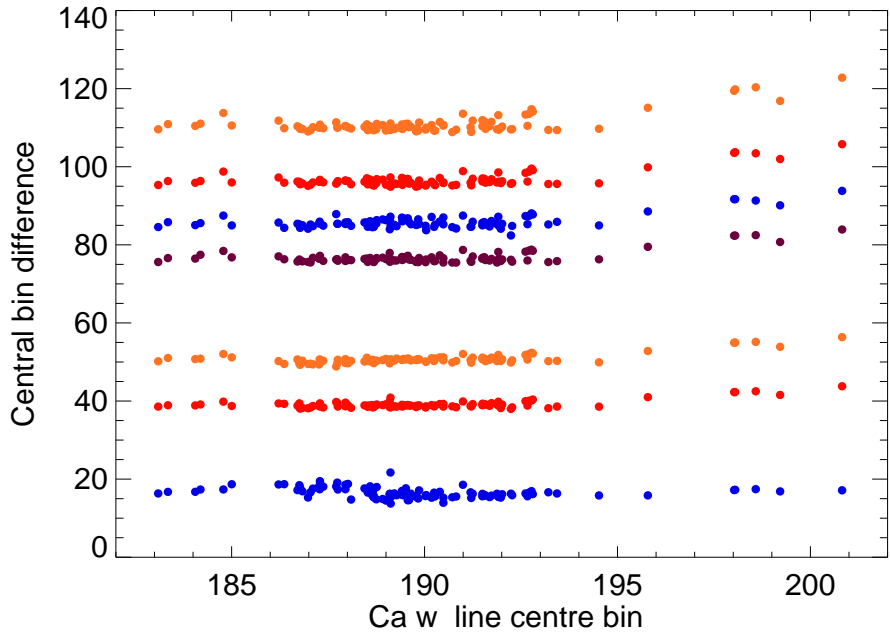}}
\caption{Distances (in bins) between bin zero and various \caxix\ lines and \caxviii\ satellites for 102 flares recorded over a wide range of offsets by the BCS before 01~January 1987. Points are colour-coded for lines observed in channel~1 -- blue: line $w$ and $d13$; red: $w$ and $x$; orange: $w$ and $y$; violet: $w$ and $q$; blue: $w$ and $r$; red: $w$ and $k$; orange: $w$ and $z$ (blended with $j$). (Line notation is from \cite{gab72}.)} \label{BCS_ch1_relative_line_posns_102flares}
\end{figure}

A further measure of the of crystal curvature uniformity for channel~1 is shown in Figure~\ref{BCS_ch1_relative_line_posns_102flares}. This shows the distance in bins of the \caxix\ line $w$ and other prominent lines from bin zero of channel~1 for flares covering a wide range of angular offsets in the dispersion plane. The line separations are close to constant, with a small deviation of the $z$-line for flares off-axis by $> 3$ arcmin in the $-x$-direction (see Table~\ref{bin_no_displacements}).

The present BCS analysis software uses the SolarSoft (SSW) package of routines which are available to users from NASA web sites. New routines for displaying BCS spectra that are in preparation and testing currently use linear relations between bin numbers and wavelength, calibrated using in-orbit fits to spectra in flare early decay stages using prominent lines with known wavelengths. The crystal radii implied by the fits differ from those of Figure~\ref{crystal_bend_radii_APThesis} by up to 30\,\% for reasons not yet established.

The translation of angular shifts in the BCS dispersion direction to wavelength, for flares offset from the BCS boresight in the dispersion plane, can be determined from the 06~November 1980 flare data. Generally, particular active regions were chosen on a daily basis during the SMM operations, so flare offsets were small as the flare was likely to be near the BCS boresight.  For such circumstances, Table~\ref{bin_no_displacements} gives the shift in BCS spectra occurring as a result of an angular offset of a flare from the boresight found from the 06~November 1980 flare. The entries in Table~\ref{bin_no_displacements} are from J.L. Lemen (personal communication, 1981) but have been verified for boresight (zero offset) values for this work. Offsets in the solar East ($-x$) direction result in a displacement to greater apparent wavelengths. For channel~1, spectra are displaced by approximately 2.57~bin~arcmin$^{-1}$, for channel~4 by 1.86~bin~arcmin$^{-1}$, and for channel~7 by 3.9~bin~arcmin$^{-1}$.

\begin{table}
\caption{Line bin numbers as a function of flare offsets}
\label{bin_no_displacements}
\begin{tabular}{llccccccc}
\hline                   
Chan.   & Line & \multicolumn{7}{c}{Offset [arcmin] in the $+x$-direction  } \\
No.     &       & +3.0  &  +2.0 & +1.0  & 0.0   & -1.0  & -2.0  & -3.0  \\
\hline
1 & \caxix\ $w$ & 182.4 & 185.0 & 187.5 & 190.1 & 192.7 & 195.2 & 197.8 \\
4 & \fexxv\ $w$ & 193.4 & 195.3 & 197.1 & 199.0 & 200.9 & 202.7 & 204.6 \\
7 & \fexxv\ $w$ & 31    & 35    & 39    & 43    & 47    & 51    & 55    \\

  \hline
\end{tabular}
\end{table}

\subsubsection{BCS Instrument Sensitivity}
\smallskip
\noindent ($a$) {\em Comparison of Spectra from Channels 5, 6, and 7 with Channel 4}

\begin{figure}
\centerline{\includegraphics[width=0.6\textwidth,clip=,angle=0]{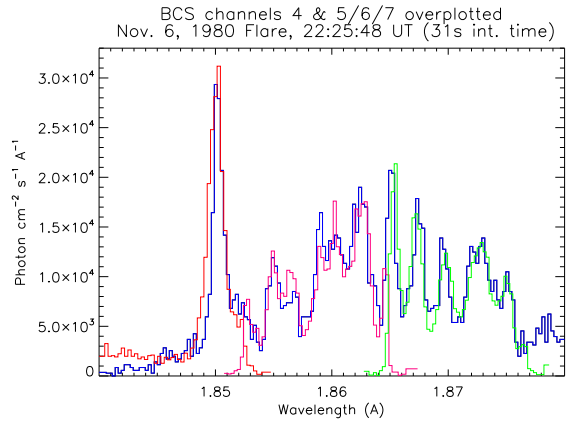}}
\caption{BCS spectra from channel~4 (blue) compared with spectra from channels 5 (green), 6 (red), 7 (orange) for a 45-second integration time starting at 22:24:07~UT during the 06~November 1980 (SOL1980-11-06T22:27: flare along BCS boresight), with counts bin$^{-1}$ converted to absolute fluxes [photon cm$^{-2}$ s$^{-1}$] using effective areas and bin widths from Section~3. The range covered includes the \fexxv\ line $w$ (1.850~\AA) and other \fexxv\ lines and dielectronic satellites of \fexxiv\ and \fexxiii. }
\label{BCS_Nov6_1980-chan4_567_sp}
\end{figure}

Opportunities for cross-calibration of BCS channels during the mission were provided by flare spectra obtained from channel~4, which observed the group of \fexxv\ lines and associated satellites in the 1.850\,--\,1.875~\AA\ range, and from channels 5, 6, and 7, which observed the same lines. Absolute spectra from all these channels, derived from effective areas (Table~\ref{BCS_eff_areas}) and bin widths, should agree if the effective areas and bin widths [\AA] are correct. This was checked for various spectra, in particular the 06~November 1980 observations discussed earlier (Sections~\ref{sec:Wavelength_Response}, \ref{sec:wave_cal}). Figure~\ref{BCS_Nov6_1980-chan4_567_sp} shows spectra on an absolute scale from channels 4, 5, 6, and 7 for a 31-second integration time at 22:27:04~UT when the flare emission was located at zero $x$-displacement (see Figure~\ref{Nov6_1980_lc_spch1}); there are 10\,\% differences or less for most of the line features observed in the range, with slightly larger amounts for the \fexxv\ $y$ (1.8594~\AA) and \fexxiv\ satellite $t$ (1.8575~\AA) lines. The effective areas of Table~\ref{BCS_eff_areas} are thus confirmed.

\medskip

\noindent ($b$) {\em Loss of the BCS Heat Shield Second Surface Mirror}

\begin{figure}
\centerline{\includegraphics[width=0.7\textwidth,clip=,angle=0]{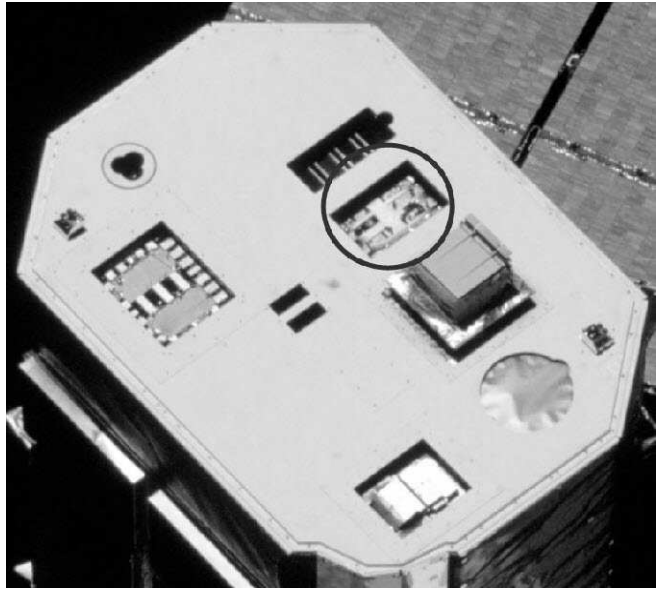}}
\caption{The front panel of the SMM spacecraft at the time of the {\it Space Shuttle} Repair Mission in April 1984, showing the BCS aperture (circled in red). The loss of the Second Surface Mirror is evident, as is the damage to the HXIS heat shield (large aperture to the left in the image). The dimensions of the front face are approximately 1.1~m~$\times$~1.6~m. Photograph courtesy of NASA. }\label{SMM_front_panel}
\end{figure}

Photographs of the Sun-facing front panel of the SMM taken by the astronauts of the SMM Repair Mission in April 1984 revealed that the BCS heat shield had suffered severe damage (Figure~\ref{SMM_front_panel}). The Teflon Second Surface Mirror (SSM) had disintegrated over all eight apertures. Some residual fragments remained attached to the side of the apertures of channels 1, 3, and 7. These extended perpendicular to the instrument face. About 5\,\% of the material over one end of channel~7 remained in place. In all cases, the Kapton thermal membrane was intact, albeit buckled.

The loss of the SSM was almost certainly the result of radiation damage by solar extreme ultraviolet (EUV) and X-ray emissions. \cite{mos05} report serious mechanical degradation of Teflon thermal wrap recovered from the {\em Hubble Space Telescope}\/ after eight years in orbit. Laboratory tests in which similar Teflon films were exposed to EUV, soft X-rays, and thermal cycling confirmed that the combination was highly damaging, with rapid and severe degradation occurring. In the case of SMM, the relatively low orbit may have resulted in residual atmospheric oxygen exacerbating the damage. A problem with plasma ingress experienced by the FCS detector system suggests that this was the case. Interaction with atomic or ionic oxygen was also identified as a possible source of discolouring of the SMM front-side white paint observed during the Repair Mission (J.B. Gurman, personal communication, 2016).

The consequences for the BCS would have occurred in two phases. Partial disintegration of the SSM in front of a given spectrometer would have introduced a position-dependent (and hence wavelength-dependent) variation in the instrument sensitivity along the dispersion axis. This would have resulted in corresponding intensity variations in the spectra recorded. The spectra would thus have been compromised during the period of mechanical collapse. However, once the SSM had disintegrated completely, the loss of the absorbing film would have led to an increased spectrometer sensitivity by a factor 1.9 for channel~1, and factors 1.12\,--\,1.15 for the remaining channels (see Table~\ref{BCS_eff_areas}). This assumes that the loss did not compromise the performance of the BCS collimator. P.~Sheather (personal communication, 2016) calculates that as a result of the SSM loss, the collimator front-grid temperatures would have increased by no more than $2^\circ$\,C relative to the design value of 20$^\circ$\,C. The collimator transmission and sidelobes would thus have been unaffected.

The photographs of the SMM front face taken by the Shuttle astronauts suggest that the HXIS instrument also suffered the damage to its Teflon heat shield. Evidence that the instrument was experiencing thermal problems began to emerge in around July 1980 (J.B. Gurman, personal communication, 2016). It may be relevant that in July 1980 a feature with no known solar explanation appeared at the low-wavelength end of the BCS channel~1. The implication is that partial loss of the HXIS and BCS SSM began in that period.

\begin{figure}
\centerline{\includegraphics[width=0.9\textwidth,clip=]{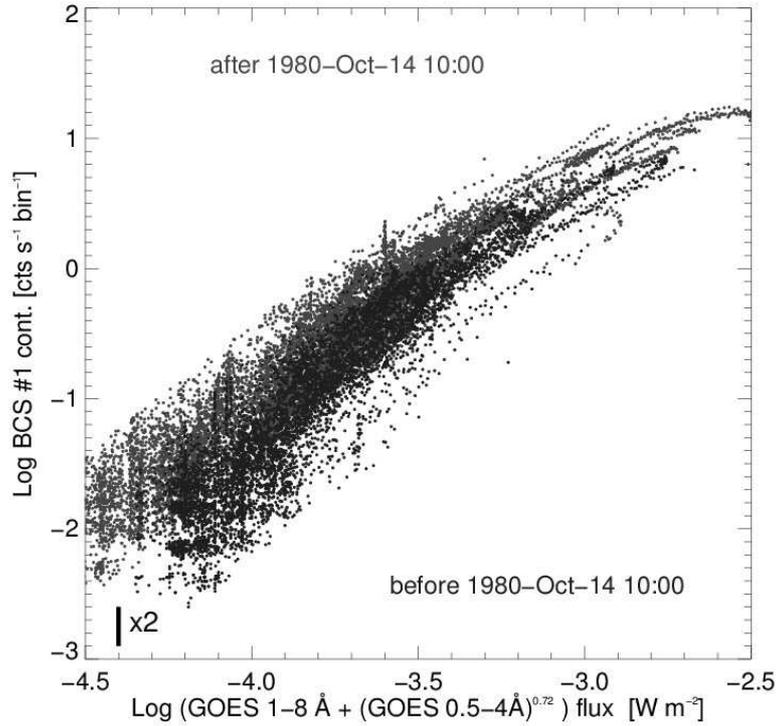}}
\caption{ Flux in a line-free portion of Channel~1 (3.168\,--\,3.171~\AA) shown on a log--log plot {\em versus} GOES fluxes in the form GOES (1\,--\,8~\AA) + GOES (0.5\,--\,4~\AA)$^{0.72}$ for flares before (dark grey) and after (light grey)  14~October 1980 (10:00~UT). The scatter is due to the lower BCS collimator transmission for flares occurring off-axis. The upward shift between the dark and light grey points by approximately a factor~2 (indicated by the black vertical bar, lower left) is apparent, and is attributed a sensitivity increase through the loss of the Second Surface Mirror (SSM). [Journal version will be in colour.]}\label{BCS_GOES_flux}
\end{figure}

To examine when the complete loss of the BCS SSM occurred, the continuum flux in channel~1 [photon counts s$^{-1}$~bin$^{-1}$] in a narrow line-free range (3.168\,--\,3.171~\AA) was plotted against the total flux in one or other of the GOES channels. The continuum flux, calculated from pre-launch effective areas (Table~\ref{BCS_eff_areas}), should show a factor-of-two relative increase following the SSM failure. Empirically it was found that plots of the channel~1 continuum against a combination of the two GOES channels, {\em viz.} GOES (1\,--\,8~\AA) + GOES (0.5\,--\,4~\AA)$^{0.72}$, gave least scatter. The remaining scatter results in part from flares that lay off the BCS boresight, and whose emission was hence reduced by the collimator response. There is evidence that the relationship between the BCS and GOES fluxes changed around 14~October 1980. Figure~\ref{BCS_GOES_flux} shows results for flares before (blue) and after (red) this time. It is concluded that the displacement of approximately a factor of 1.9 between the red and blue points represents the loss of the channel~1 SSM. This may have occurred over a period of time, though possibly the SSM disintegration was catastrophic between 13~October and 14~October. It may be relevant that one of the most intense flares observed by the SMM instruments (GOES class X3) occurred on 14~October (peak 06:11: SOL1980-10-14T06:11), raising the prospect that it may have delivered the {\em coup de grace}.

Direct and continuous exposure to the radiation from the highly active Sun, exacerbated by residual atmospheric oxygen, proved too severe for the BCS (and HXIS) heat-shield design. It is ironic that Teflon film was selected for its supposed radiation-hardness, and it was the recommended material for both the BCS and HXIS front membranes. The use of a gridded shield, as for the FCS, would have overcome the problem but would have been difficult to accommodate within the spacecraft space limitations. However, it offers a solution for future instruments.

\subsubsection{BCS Alignment and Collimator Response}

The BCS boresight was compared with HXIS and FCS boresights from two flares (14~August and 06~November) in 1980 \citep{wag86} when the SMM spacecraft made raster scans while the flares were in progress. With respect to the SMM {\em Fine Pointing Sun Sensor} axis (FPSS), the spacecraft raster scans indicated that the BCS boresight was offset $+29 \pm 15$~arcsec in $x$ and $+42 \pm 20$~arcsec in $y$ with an accuracy of about $\pm 30$~arcsec. The same data allowed the collimator transmission to be checked against the pre-launch measurements: this was confirmed to be $6.0 ^{+.5}_{0.0}$~arcmins (FWHM). The small misalignment of the BCS with respect to the FPSS boresight was well within specification and resulted in very small corrections ($< 15$\,\%) for collimator transmission.

\begin{figure}
\centerline{\includegraphics[width=0.7\textwidth,clip=,angle=0]{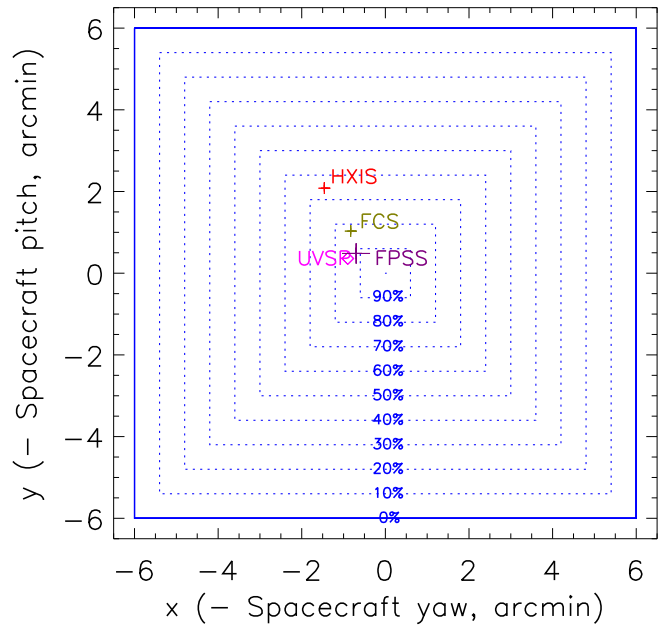}}
\caption{BCS collimator response in spacecraft coordinates ($x$, $y$) shown as squares (dashed lines) with collimator response in percentages of the peak (central) value. Also shown are the boresight positions (with estimated uncertainties) of the spacecraft FPSS (purple), the FCS (dark green: see text) and HXIS (red) instruments. The UVSP boresight is indicated (magenta diamond) but no uncertainties are available. The boresight positions are based on 1980 flare data \citep{wag86}. The BCS dispersion is in the E\,--\,W ($\pm x$) direction. }
\label{bcs_coll_response}
\end{figure}

The angular response of the BCS collimator is shaped like a square pyramid, with FWHM equal to 6.0~arcmin and with maximum response at the BCS boresight equal to $33 \pm 3$\,\% (measured: Section~\ref{collimator}). This is illustrated in terms of the standard solar coordinate system in Figure~\ref{bcs_coll_response}. The spacecraft boresight with respect to this response function is shown, as are those for other SMM instruments (FCS, HXIS, UVSP) for flares observed in the 1980 period of operations using data of \cite{wag86}. It should be noted that the FCS instrument had a fine collimator (angular resolution of $\approx 14$~arcsec) with a field of view that could be rastered over a range of $7 \times 7$~arcmin in 5~arcsec steps. The point marked FCS in Figure~\ref{bcs_coll_response} shows the centre of the FCS raster range.

\subsection{Detector Issues}
\label{det_iss}

Before launch, a small dead region, called the ``notch'', was found to exist in the central response of each detector position output. This corresponded to an offset of the two halves of the output, and a loss of sensitivity of the adjacent position bins. It was an artifact of the flight electronics design, and it had not been encountered with the laboratory test electronics, nor with the prototype spectrometer flown previously as an Aerobee rocket payload \citep{rap77}. The problem was discovered too late in the delivery programme to be rectified, but it was compensated for in the data-processing software. A tailored correction was applied to each channel, removing the discontinuity between the two halves of the spectral output, and adjusting the count rates in the central and adjacent bins by comparing them with adjacent bins. This worked fairly well in cases where the spectral intensity was slowly changing across the region, but less so when it coincided with the edge of a spectral line.  The output of all eight detectors obtained during passages through the South Atlantic Anomaly at a time when no flare was occurring revealed residual features related to the notch, indicating that the ``wings'' extended over several adjacent bins. An additional software correction is being developed for SolarSoft. In the meantime, spectral features within $\pm 5$ bins of the detector centre should be treated with caution.

The BCS Fe-channel detectors 2 and 8 suffered slowly increasing gain attributed to low rates of gas leakage, probably as a result of loss of the hermetic ``O''-ring seal due to vibration during launch. The former became inoperable after approximately seven months, whilst the latter operated for over two years. Pulse-height analyzer data from the remaining detectors showed no significant changes over the duration of the mission.

At detector count-rates of $\gtrsim 5000$~s$^{-1}$ expansion of the detector position gain (and hence spectral dispersion) occurred. A count rate of $\approx 5000$~s$^{-1}$ in channel~1 led to an increase of the \caxix\ $w - z$ line separation by approximately 1.5~bins. This was due to pulse pile-up in the analogue electronics and was a known performance limitation. Above $\approx 5000$~s$^{-1}$, the measured count rate was also increasingly an underestimate owing to the operation of the digital electronics, as described in Section~\ref{anal_elec}. In the case of especially intense flares, notably the X10 flare on 24~April 1984 (SOL1984-04-24T00:23), in which count rates reached values as high as 8000\,--\,9000~s$^{-1}$, severe spectral distortions occurred. Occasions when the spectral distortion partially recovered at extremely high count rates are difficult to explain, and these require further investigation. The lesson for future missions is to test and characterize the upper dynamic range of the spectrometers pre-launch.

Flare BCS spectra have been widely used to determine electron temperatures and emission measures from line ratios and line intensities ({\em e.g.} Lemen {\em et al.}, 1984), so detector effects such as the notch and pulse-pileup are of some importance. An early attempt to analyze BCS channel~1 spectra using \caxix\ and \caxviii\ lines \citep{bel82b} found a significant discrepancy between the observed \caxix\ line $y$ (3.1925~\AA) intensity for the X1.4 flare on 21~May 1980 (SOL1980-05-21T21:07) and that calculated from theory. This was attributed to an unknown instrumental effect. However, flare spectra from the {\em P78-1}/SOLEX instrument \citep{see84} and the Alcator tokamak \citep{rice14} confirm the BCS observed intensity, so the error appears to lie with the calculations and is not a detector effect. Another possible problem with the \caxviii\ satellite $q$, which is important for diagnosing departures from ionization equilibrium, is acknowledged by \cite{bel82b} to be due to a blend with an Ar~{\sc xvii} line and again is not due to detector effects.

\subsection{Crystal Fluorescence}
\label{cryst_fluor}

\begin{figure}
\centerline{\includegraphics[width=0.9\textwidth,clip=,angle=0]{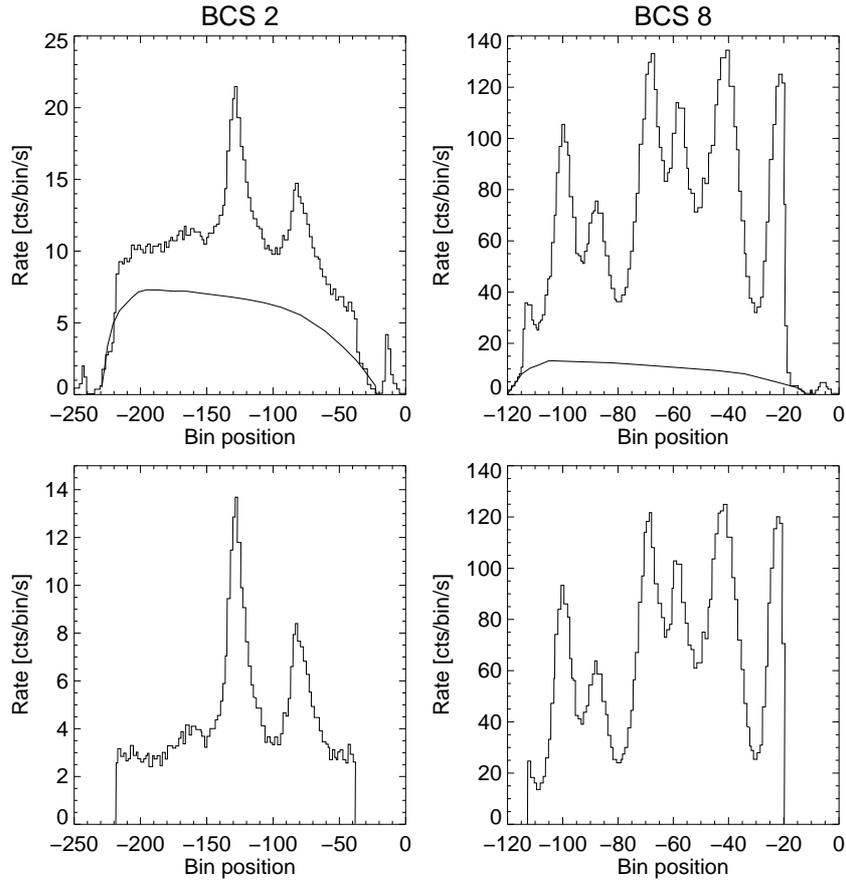}}
\caption{Analysis by Parmar (1981) of spectra in BCS channels 2 (left panels) and 8 (right panels) in which fluorescence emission, modeled with a Monte Carlo routine (smooth curves in upper panels), is removed to give the corrected spectra in the lower panels.\label{Ge_fluor_APThesis}}
\end{figure}

The BCS was subject to fluorescence of the crystal material caused by solar X-rays. This formed a continuous background, or curved ``pedestal'', for spectra in the Fe channels. The effect appears to be very small in channel~1. As shown in an analysis by \cite{parm81}, this
pedestal reduces the signal-to-noise ratio for faint spectral features, affects estimates of the solar continuum flux, and distorts spectral line profiles. Pulse-height analysis (PHA) data of flare spectra compared with $^{55}$Fe calibration data show a peak at about 10~keV (the upper limit of the PHA) confirming that the pedestal counts correspond to 9.9~keV emission due to germanium fluorescence.  \cite{parm81} wrote a Monte-Carlo-based geometric model code of each BCS channel in order to compute a theoretical pedestal shape and relative intensity. This included i) a test for photons absorbed by the crystal holder; ii) a small ($2-3$\,\%) correction to take into account the geometry of the crystal curvature; iii) a correction for the angular transmission of the detector honeycomb collimator; iv) a position correction for absorption in the detector gas (mean free path of 9.9~keV X-rays is 8~mm); and v) an estimate of the germanium fluorescence flux, measured at each end of the detector, beyond the limit of the solar spectrum. Figure~\ref{Ge_fluor_APThesis} shows the analysis by \cite{parm81} for BCS channels~2 and 6 where the modelled fluorescence (continuous curves in the upper two panels) was removed to produce the corrected spectra in the lower panels.

Commandable lower and upper thresholds on the analogue electronics single channel analyzers would have allowed ``tuning'' to reduce the sensitivity to the 9.9~keV fluorescence radiation, as was in fact done for the {\em Rentgenovsky Spektrometr s Izognutymi Kristalami} (RESIK) solar crystal spectrometer on the {\em CORONAS-F} spacecraft \citep{jsyl05}.

\subsection{Background Count Rates}

The BCS detector cosmic-ray particle background were determined from interpolation of rates recorded in the low- and high-wavelength non-solar slots (which are correlated). The rates varied around the orbit with count rates in the range $\approx$ 0.001\,--\,0.03~bin$^{-1}$~s$^{-1}$ (channel~1) and $\approx$ 0.01\,--\,0.08~bin$^{-1}$~s$^{-1}$ (channels~2 to 8). Since the events were distributed throughout the detector lengths, the rates per digital bin were generally insignificant relative to the X-ray continuum and line count rates from flares greater than class C flares and stronger, except for channels~3 and 8. Given the weak signals detected in the latter channels, it was necessary to subtract an estimate of the cosmic-ray rate derived from the counts in the end regions of the detectors not illuminated by solar X-rays. \cite{ben86} reported the use of cosmic-ray background rates derived from the FCS detectors to estimate BCS detector backgrounds, but concluded the method offered no advantage over the use of the BCS detector end regions. Occasionally, as a result of imperfections in the geographic mask used to turn off detectors during passage through the South Atlantic Anomaly, count rates as high as 100~~bin$^{-1}$~s$^{-1}$ were observed.

\section{BCS Spectra and Analysis}
\label{sp_anal}

BCS data files, known as bda files, are currently in the public domain and are available from the NASA ftp site \newline \texttt{ftp://umbra.nascom.nasa.gov/smm/xrp/data/bda/}. \newline They are arranged by the year of the data (1980 and 1984\,--\,1989), and named by the date (format yymmdd) and time (UT) of the start of the data range (format hhmm); thus the first file of the 1980 data files, from 28~February 1980 with start time 23:32~UT, is \texttt{bda800228.2332}.

At present, the spectra may be analyzed using IDL routines available in the {\sf SSWIDL} package \citep{fre98}. A software guide for all the IDL routines used to analyze the XRP (BCS and FCS) data is available from the ftp site \newline \texttt{ftp://umbra.nascom.nasa.gov/smm/xrp/documents}. \newline The BCS routines operate on the bda files using pre-launch values of BCS effective areas and dispersion (Tables~\ref{BCS_eff_areas} and \ref{wave_cal}) to produce flare spectra with absolute spectral irradiance units [photons cm$^{-2}$~s$^{-1}$~\AA$^{-1}$]. The analysis techniques that are now being developed will allow much more detailed analysis. First, the dispersion non-uniformities (Figure~\ref{crystal_bend_radii_APThesis}) and the stretching over the SMM lifetime (Section~\ref{sec:det_linear}) have been estimated and can be allowed for. The amounts of each will be incorporated into rewritten IDL routines. Preliminary versions of routines analyzing channels~1 and 4 (including the diagnostically important \caxix\ and \fexxv\ lines and associated dielectronic satellites) that allow for spectrometer non-linearities (Figure~\ref{wvl_bin_plot}) will be made available as soon as possible, while in the longer term, programs incorporating crystal non-uniformities and the spectral stretching will be written. Secondly, refined estimates of the wavelength resolution including the wavelength dependence (Section~\ref{sec:wave_res}, Table~\ref{BCS_wave_res}) for each channel now allow the raw BCS data to be deconvolved. Preliminary programs can now produce flare spectra with much enhanced clarity of the satellite structure in the \caxix\ and \fexxv\ lines of channels~1 and 4. This enables temperature to be determined more precisely, and offer better diagnostic capability for examining the nature of soft X-ray flare plasmas.

\begin{figure}
\centerline{\includegraphics[width=0.95\textwidth,clip=,angle=0]{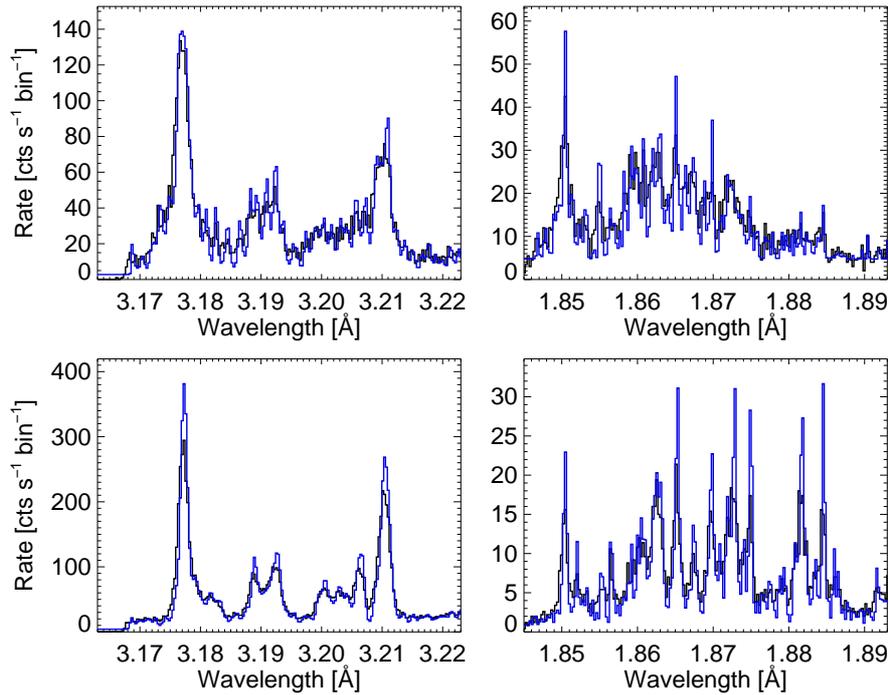}}
\caption{Channel~1 (upper left) and channel~4 (upper right) spectra are shown for the rising phase (09:17\,--\,09:18~UT) of a flare on 10~April 1980 (SOL1980-04-10T09:22). The black histograms are the raw data and the blue histograms are the deconvolved data. The flare rise phase coincides with a hard X-ray peak; broadened lines and extra emission on the short-wavelength side (``blue shifts'') of all lines are evident. Equivalent spectra are shown (lower left and lower right) for the flare's post-peak (09:27\,--\,09:28~UT), when the blue shifts are no longer apparent. } \label{bcs_spectra_for_selected_times}
\end{figure}

As an illustration, Figure~\ref{bcs_spectra_for_selected_times} shows channel~1 and 4 spectra at two stages during the flare of 10~April 1980 (SOL1980-04-10T09:22). The spectra are in ``raw'' units of photon counts s$^{-1}$~bin$^{-1}$, {\em i.e.} the effective-area factor has not been applied. The dispersion is that determined from pre-launch values for 1980 (Table~\ref{wave_cal}). The raw spectrum (in black) shows spectral-line features that are wider than those in the deconvolved spectra (in blue). As the counts are conserved, all lines have peak spectral irradiances that are substantially larger. The complex satellite structure in channel~4 particularly leads to a piling up of emission, but in the deconvolved spectra the satellite interline spacing goes down nearly to the background level defined by the region to the short-wavelength side of line $w$.

\begin{figure}
\centerline{\includegraphics[width=0.9\textwidth,clip=,angle=0]{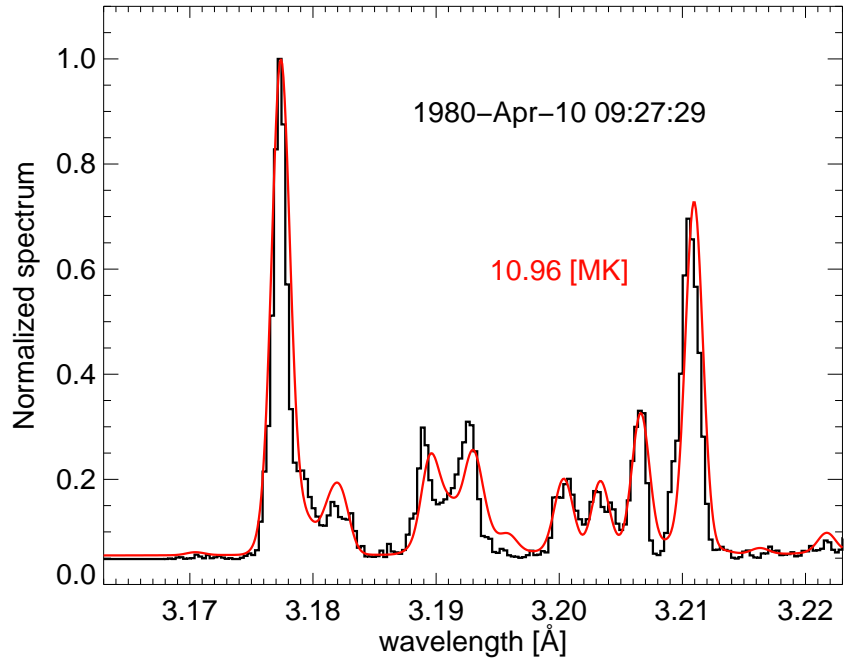}}
\caption{BCS spectrum during the decay of the flare of 10~April 1980 (SOL1980-04-10T09:22) fitted with the synthetic spectrum program described in the text for an electron temperature of 10.96~MK. Spectrum is normalized to the \caxix\ $w$ line at 3.177~\AA.} \label{CaXIX_sp_fit_Apr10_1980}
\end{figure}

It is instructive to see how the improved resolution of BCS channel~1 flare spectra resulting from deconvolution now agrees with theoretical \caxix\ spectra. The latter have been generated from a combination of data from \cite{bel82b}, the {\sf CHIANTI} atomic data and software package, and runs of the Cowan Hartree-Fock program (for high-$n$ satellites not included in {\sf CHIANTI}) in a program that has been written for a work in preparation. Figure~\ref{CaXIX_sp_fit_Apr10_1980} shows a spectrum from the 10~April 1980 flare with raw spectrum (normalized to the peak of the \caxix\ line $w$) and the new theoretical spectrum. Apart from slight discrepancies in line wavelengths (some from \cite{see89}), the agreement is excellent. There is very close agreement of the observed background with the theoretical continuum, calculated from {\sf CHIANTI} functions, implying a very small amount of crystal fluorescence in channel~1 (Section~\ref{cryst_fluor}).

\section{Summary and Conclusions}
\label{sec:summary}

In this work, we have described the instrumentation of the SMM/{\em Bent Crystal Spectrometer} in more detail than in previous publications. Bent-crystal technology remains highly relevant for X-ray spectroscopy of solar flares. The temporal resolution from spectrometers built with curved crystals and position-sensitive detectors greatly exceeds that of flat, scanning-crystal spectrometers. At present, they are still to be preferred over the newer-technology micro-calorimeters that need liquid helium cryogenics and to date have only been demonstrated for laboratory uses and non-solar X-ray sources. Future space missions that will have bent crystal spectrometers onboard include the Sun-orbiting {\em Interhelioprobe}\/ spacecraft with the ChemiX spectrometer with ``Dopplerometer'' techniques that will allow flare plasma Doppler shifts to be distinguished from emitting plasma transverse motions \citep{jsyl15,sia16}. Charge-coupled devices (CCDs) will replace the position-sensitive proportional counters used in the SMM/BCS, eliminating a number of instrumental artefacts described here.

Important results from this work include new estimates of BCS effective areas (Table~\ref{BCS_eff_areas}) enabling spectra in absolute units to be calculated and wavelength resolution (Table~\ref{BCS_wave_res}), both from pre-launch measurements, which are given here for the first time. In-orbit measurements are described, based on particular BCS flare observations. The spacecraft scans during the 06~November 1980 flare have been especially useful in wavelength calibration (bin-to-wavelength conversion for flares offset from the instrument boresight in the dispersion plane: see Table~\ref{wave_cal}) and measuring crystal curvature anomalies (Figure~\ref{crystal_bend_radii_APThesis}). Departures of spectrometer linearity are given for channels~1 and 4 (which have relatively large wavelength ranges) in Figure~\ref{fitted_ch1_ch4_lines_vs_time}, indicating only slight departures but ones that can easily be allowed for in analysis software. The detectors' position gain and spectral shift as the instrument aged over the spacecraft lifetime are shown in Figure~\ref{fitted_ch1_ch4_lines_vs_time}, which indicate a spectral linear stretch associated with the evolution of the product of the detector anode resistance and end capacitance in some channels. Comparison of channel 4 flare spectra with those of channels~5, 6, and 7 covering the same range of \fexxv\ and \fexxiv\ lines (Figure~\ref{BCS_Nov6_1980-chan4_567_sp}) shows that the effective areas measured pre-launch are very satisfactorily confirmed for these channels at least. Photographs of the SMM front panel by the Space Shuttle astronauts during the STS-41C Repair Mission revealed the disintegration of the heat shield Second Surface Mirror (SSM). This appears not to have had a significant impact on the instrument (the temperature rise of the BCS is estimated to be well within specifications) other than a factor-of-two increase in the sensitivity of channel~1 once the disintegration was complete. Comparison of a line-free region of flare spectra with GOES emission over the 1980 period indicates that there was indeed an increase in sensitivity around mid-October. The detector notch was an unfortunate design anomaly, although its effect can be removed to some extent by analysis software. Crystal fluorescence is significant for channels~2 to 8. Modelling the wavelength dependence requires time-consuming Monte Carlo techniques, which is not a reasonable option for all spectra during all events. However, the fluorescence in channel~1 seems to be very small, as seen from the agreement of the background level with the continuum in modelled spectra (Figure~\ref{CaXIX_sp_fit_Apr10_1980}).

A number of recommendations can be made with the experience of the BCS over its lifetime. Achieving uniform and stable crystal curvatures is a core requirement. The performance of the BCS in this respect could be improved upon. Crystals in preparation for the {\em ChemiX} spectrometer \citep{jsyl15,sia16} are glued to mandrels and laboratory tests show encouraging results. To determine lifetime effects prior to launch, such as detector gain changes, stronger X-ray sources than those used are clearly desirable. The disintegration of the SSM part of the heat shield, apparently during the first year of operation (although only detected in 1984), would have been avoided with the use of a gridded heat shield as was used for the XRP/FCS instrument. The decision to use Teflon as a heat shield material proved unwise, and its use under high irradiation should be ruled out in future. Fluorescence needs to be considered in the choice of crystal used. The provision of tunable single-channel analyzers in the analogue electronics chains could have helped reduce its effect. The occurrence of detector anomalies at very high count rates (as happened during the April 1984 flares) underscores the need to characterize the upper dynamic range of detectors before launch.

With new assessments of the BCS wavelength resolution, the archived flare data can now be re-analyzed with de-convolution techniques. Using Interactive Data Language (IDL) routines unavailable during the SMM period, high-quality spectral data can be achieved with temporal plots that reveal the changing spectral-line profiles at the impulsive phases of flares, as has been illustrated in Section~\ref{sp_anal} (Figure~\ref{bcs_spectra_for_selected_times}). The close agreement of channel~1 flare spectra with modelled spectra (Figure~\ref{CaXIX_sp_fit_Apr10_1980}) is a clear indication of the worth of BCS spectra. It should be noted that the 1980 and 1984 periods had much higher levels of solar activity than in more recent years, so that BCS data may be unsurpassed for some years to come in the investigation of the solar-flare phenomenon.


\begin{acks}
The SMM X-ray Polychromator was designed, tested, and constructed by Lockheed Missiles and Space Co. (LMSC; PI L.W. Acton), University College London Mullard Space Science Laboratory (MSSL; PI J.L. Culhane), and the UK Rutherford Appleton Laboratory (RAL; PI A.H. Gabriel). The BCS instrument was primarily the responsibility of MSSL (Project Scientist C.G. Rapley). The crystals and their mounts were provided by RAL (Project Engineer R.F. Turner) and the digital control and data processing electronics, including the microprocessors, by LMSC (Project Scientist C.J. Wolfson). We are grateful to all involved, and to the astronauts of the Space Shuttle Repair Mission of April 1984, notably the late Francis R. Scobee who died in the {\em Challenger}\/ disaster of 1986. Without the achievement of the Repair Mission astronauts Nelson and VanHoften, no BCS data after November 1980 would have been available. We also thank Peter Waggett, Arvind Parmar, and Robert Bentley whose PhD theses from the early 1980s record efforts and results which have been invaluable in the writing of this paper. We are grateful to Joseph B. Gurman (NASA/GSFC) for his assistance in obtaining imagery from STS-41C, and to Craig Theobald (UCL Mullard Space Science Laboratory) for his support in locating the original mechanical drawings of the BCS. We are particularly grateful to Peter Sheather for help with assessing the likely impact of the disintegrated SSM on the temperature of the BCS instrument. Dominic Zarro and James~R. Lemen provided the user-friendly software which has allowed BCS data to be extensively analyzed and interpreted over the past few years, and Amy Forinash at NASA/GSFC maintains the XRP database including the BCS data files and documentation. We have also benefited from the considerable help of Barbara Sylwester, Zaneta Szaforz, Jaroslav Baka{\l}a, and Anna K\k{e}pa at SRC in Wroc{\l}aw with calculations, figures, and analysis and for creating the web catalogue of spectra. We acknowledge the support of Polish National Science Centre grants 2013/11/B/ST9/00234 and 2015/19/B/ST9/02826. J.~Sylwester is grateful for the support from the International Space Science Institute in Bern, Team No. 276 (``Non-Equilibrium Processes in the Solar Corona and their Connection to the Solar Wind'').
\end{acks}

\section*{Disclosure of Potential Conflicts of Interest}

The authors declare that they have no conflicts of interest.

%
\bibliographystyle{spr-mp-sola}
\bibliography{RESIK}

\begin{thebibliography}{33}
\ifx\bisbn     \undefined \def\bisbn  #1{ISBN #1}\fi
\ifx\binits    \undefined \def\binits#1{#1}\fi
\ifx\bauthor   \undefined \def\bauthor#1{#1}\fi
\ifx\batitle   \undefined \def\batitle#1{#1}\fi
\ifx\bjtitle   \undefined \def\bjtitle#1{\textit{#1}}\fi
\ifx\bvolume   \undefined \def\bvolume#1{\textbf{#1}}\fi
\ifx\byear     \undefined \def\byear#1{#1}\fi
\ifx\bissue    \undefined \def\bissue#1{#1}\fi
\ifx\bfpage    \undefined \def\bfpage#1{#1}\fi
\ifx\blpage    \undefined \def\blpage #1{#1}\fi
\ifx\burl      \undefined \def\burl#1{\textsf{#1}}\fi
\ifx\href      \undefined \def\href#1#2{\textsf{#2}}\fi
\ifx\betal     \undefined \def\betal{\textit{et al.}}\fi
\ifx\bctitle   \undefined \def\bctitle#1{#1}\fi
\ifx\beditor   \undefined \def\beditor#1{#1}\fi
\ifx\bbtitle   \undefined \def\bbtitle#1{\textit{#1}}\fi
\ifx\bedition  \undefined \def\bedition#1{#1}\fi
\ifx\bseriesno \undefined \def\bseriesno#1{\textbf{#1}}\fi
\ifx\blocation \undefined \def\blocation#1{#1}\fi
\ifx\bsertitle \undefined \def\bsertitle#1{\textit{#1}}\fi
\ifx\bsnm      \undefined \def\bsnm#1{#1}\fi
\ifx\bsuffix   \undefined \def\bsuffix#1{#1}\fi
\ifx\bparticle \undefined \def\bparticle#1{#1}\fi
\ifx\barticle  \undefined \def\barticle#1{}\fi
\ifx\binstitute  \undefined \def\binstitute#1{#1}\fi
\ifx\bpublisher  \undefined \def\bpublisher#1{#1}\fi
\ifx\doiurl    \undefined
  \def\doiurl#1{\href{http://dx.doi.org/#1}{\textsf{DOI}}}\fi
\ifx\arxivurl  \undefined
  \def\arxivurl#1{\href{http://arxiv.org/abs/#1}{\textsf{arXiv}}}\fi
\ifx\adsurl    \undefined
  \def\adsurl#1{\href{http://adsabs.harvard.edu/abs/#1}{\textsf{ADS}}}\fi
\ifx\botherref \undefined \def\botherref#1{}\fi
\ifx\url       \undefined \def\url#1{\textsf{#1}}\fi
\ifx\bchapter  \undefined \def\bchapter#1{}\fi
\ifx\bbook     \undefined \def\bbook#1{}\fi
\ifx\bcomment  \undefined \def\bcomment#1{#1}\fi
\ifx\oauthor   \undefined \def\oauthor#1{#1}\fi
\ifx\citeauthoryear \undefined\def \citeauthoryear#1{#1}\fi
\ifx\endbibitem\undefined \def\endbibitem{}\fi
\ifx\bconflocation  \undefined \def\bconflocation#1{#1} \fi

\bibitem[\protect\citeauthoryear{{Acton} \textit{et~al.}}{1980}]{act80}
\begin{barticle}
\bauthor{\bsnm{{Acton}}, \binits{L.W.}},
\bauthor{\bsnm{{Finch}}, \binits{M.L.}},
\bauthor{\bsnm{{Gilbreth}}, \binits{C.W.}},
\bauthor{\bsnm{{Culhane}}, \binits{J.L.}},
\bauthor{\bsnm{{Bentley}}, \binits{R.D.}},
\bauthor{\bsnm{{Bowles}}, \binits{J.A.}},
\bauthor{\bsnm{{Guttridge}}, \binits{P.}},
\bauthor{\bsnm{{Gabriel}}, \binits{A.H.}},
\bauthor{\bsnm{{Firth}}, \binits{J.G.}},
\bauthor{\bsnm{{Hayes}}, \binits{R.W.}},
\bauthor{\bsnm{{Joki}}, \binits{E.G.}},
\bauthor{\bsnm{{Jones}}, \binits{B.B.}},
\bauthor{\bsnm{{Kent}}, \binits{B.J.}},
\bauthor{\bsnm{{Leibacher}}, \binits{J.W.}},
\bauthor{\bsnm{{Nobles}}, \binits{R.A.}},
\bauthor{\bsnm{{Patrick}}, \binits{T.J.}},
\bauthor{\bsnm{{Phillips}}, \binits{K.J.H.}},
\bauthor{\bsnm{{Rapley}}, \binits{C.G.}},
\bauthor{\bsnm{{Sheather}}, \binits{P.H.}},
\bauthor{\bsnm{{Sherman}}, \binits{J.C.}},
\bauthor{\bsnm{{Stark}}, \binits{J.P.}},
\bauthor{\bsnm{{Springer}}, \binits{L.A.}},
\bauthor{\bsnm{{Turner}}, \binits{R.F.}},
\bauthor{\bsnm{{Wolfson}}, \binits{C.J.}}:
\byear{1980},
\batitle{{The soft X-ray polychromator for the Solar Maximum Mission}}.
\bjtitle{\solphys}
\bvolume{65},
\bfpage{53}.
\doiurl{10.1007/BF00151384}.
\adsurl{1980SoPh...65...53A}.
\end{barticle}
\endbibitem

\bibitem[\protect\citeauthoryear{{Antonucci} \textit{et~al.}}{1982}]{ant82}
\begin{barticle}
\bauthor{\bsnm{{Antonucci}}, \binits{E.}},
\bauthor{\bsnm{{Gabriel}}, \binits{A.H.}},
\bauthor{\bsnm{{Acton}}, \binits{L.W.}},
\bauthor{\bsnm{{Leibacher}}, \binits{J.W.}},
\bauthor{\bsnm{{Culhane}}, \binits{J.L.}},
\bauthor{\bsnm{{Rapley}}, \binits{C.G.}},
\bauthor{\bsnm{{Doyle}}, \binits{J.G.}},
\bauthor{\bsnm{{Machado}}, \binits{M.E.}},
\bauthor{\bsnm{{Orwig}}, \binits{L.E.}}:
\byear{1982},
\batitle{{Impulsive phase of flares in soft X-ray emission}}.
\bjtitle{\solphys}
\bvolume{78},
\bfpage{107}.
\doiurl{10.1007/BF00151147}.
\adsurl{1982SoPh...78..107A}.
\end{barticle}
\endbibitem

\bibitem[\protect\citeauthoryear{{Bearden}}{1967}]{bea67}
\begin{barticle}
\bauthor{\bsnm{{Bearden}}, \binits{J.A.}}:
\byear{1967},
\batitle{{X-Ray Wavelengths}}.
\bjtitle{Rev. Mod. Phys.}
\bvolume{39},
\bfpage{78}.
\doiurl{10.1103/RevModPhys.39.78}.
\adsurl{1967RvMP...39...78B}.
\end{barticle}
\endbibitem

\bibitem[\protect\citeauthoryear{{Bely-Dubau} \textit{et~al.}}{1982}]{bel82b}
\begin{barticle}
\bauthor{\bsnm{{Bely-Dubau}}, \binits{F.}},
\bauthor{\bsnm{{Faucher}}, \binits{P.}},
\bauthor{\bsnm{{Steenman-Clark}}, \binits{L.}},
\bauthor{\bsnm{{Dubau}}, \binits{J.}},
\bauthor{\bsnm{{Loulergue}}, \binits{M.}},
\bauthor{\bsnm{{Gabriel}}, \binits{A.H.}},
\bauthor{\bsnm{{Antonucci}}, \binits{E.}},
\bauthor{\bsnm{{Volonte}}, \binits{S.}},
\bauthor{\bsnm{{Rapley}}, \binits{C.G.}}:
\byear{1982},
\batitle{{Dielectronic satellite spectra for highly-charged helium-like ions.
  VII - Calcium spectra: Theory and comparison with SMM observations}}.
\bjtitle{\mnras}
\bvolume{201},
\bfpage{1155}.
\adsurl{1982MNRAS.201.1155B}.
\end{barticle}
\endbibitem

\bibitem[\protect\citeauthoryear{{Bentley}}{1986}]{ben86}
\begin{botherref}
\oauthor{\bsnm{{Bentley}}, \binits{R.D.}}:
1986,
{Soft X-ray emission from solar flares and active regions}.
PhD thesis,
Mullard Space Science Laboratory, University College London, Holmbury St.~Mary,
  Dorking, Surrey RH5 6NT, UK.
\end{botherref}
\endbibitem

\bibitem[\protect\citeauthoryear{{Borkowski} and {Kopp}}{1968}]{bor68}
\begin{barticle}
\bauthor{\bsnm{{Borkowski}}, \binits{C.J.}},
\bauthor{\bsnm{{Kopp}}, \binits{M.K.}}:
\byear{1968},
\batitle{{New Type of Position-Sensitive Detectors of Ionizing Radiation Using
  Risetime Measurement}}.
\bjtitle{Rev. Sci. Instr.}
\bvolume{39},
\bfpage{1515}.
\doiurl{10.1063/1.1683152}.
\adsurl{1968RScI...39.1515B}.
\end{barticle}
\endbibitem

\bibitem[\protect\citeauthoryear{{Catura} \textit{et~al.}}{1974}]{cat74}
\begin{barticle}
\bauthor{\bsnm{{Catura}}, \binits{R.C.}},
\bauthor{\bsnm{{Joki}}, \binits{E.G.}},
\bauthor{\bsnm{{Bakke}}, \binits{J.C.}},
\bauthor{\bsnm{{Rapley}}, \binits{C.G.}},
\bauthor{\bsnm{{Culhane}}, \binits{J.L.}}:
\byear{1974},
\batitle{{A new type of crystal spectrometer for solar X-ray studies}}.
\bjtitle{\mnras}
\bvolume{168},
\bfpage{217}.
\adsurl{1974MNRAS.168..217C}.
\end{barticle}
\endbibitem

\bibitem[\protect\citeauthoryear{{del R{\'{\i}}o} and {Dejus}}{2004}]{xop04}
\begin{bchapter}
\bauthor{\bsnm{{del R{\'{\i}}o}}, \binits{M.S.}},
\bauthor{\bsnm{{Dejus}}, \binits{R.J.}}:
\byear{2004},
\bctitle{{XOP 2.1 - A New Version of the X-ray Optics Software Toolkit}}.
In: \beditor{\bsnm{{Warwick}}, \binits{T.}},
\beditor{\bsnm{{Arthur}}, \binits{J.}},
\beditor{\bsnm{{Padmore}}, \binits{H.A.}},
\beditor{\bsnm{{St{\"o}hr}}, \binits{J.}} (eds.)
\bbtitle{American Institute of Physics Conference Series},
\bsertitle{Am. Inst. Phys.}
\bseriesno{705},
\bfpage{784}.
\doiurl{10.1063/1.1757913}.
\adsurl{2004AIPC..705..784D}.
\end{bchapter}
\endbibitem

\bibitem[\protect\citeauthoryear{{Doschek}, {Kreplin}, and
  {Feldman}}{1979}]{dos79}
\begin{barticle}
\bauthor{\bsnm{{Doschek}}, \binits{G.A.}},
\bauthor{\bsnm{{Kreplin}}, \binits{R.W.}},
\bauthor{\bsnm{{Feldman}}, \binits{U.}}:
\byear{1979},
\batitle{{High-resolution solar flare X-ray spectra}}.
\bjtitle{\apjl}
\bvolume{233},
\bfpage{L157}.
\doiurl{10.1086/183096}.
\adsurl{1979ApJ...233L.157D}.
\end{barticle}
\endbibitem

\bibitem[\protect\citeauthoryear{{Doschek}, {Meekins}, and
  {Cowan}}{1972}]{dos72}
\begin{barticle}
\bauthor{\bsnm{{Doschek}}, \binits{G.A.}},
\bauthor{\bsnm{{Meekins}}, \binits{J.F.}},
\bauthor{\bsnm{{Cowan}}, \binits{R.D.}}:
\byear{1972},
\batitle{{Further Iron-Line Observations during Solar Flares}}.
\bjtitle{\apj}
\bvolume{177},
\bfpage{261}.
\doiurl{10.1086/151704}.
\adsurl{1972ApJ...177..261D}.
\end{barticle}
\endbibitem

\bibitem[\protect\citeauthoryear{{Doschek} \textit{et~al.}}{1971}]{dos71}
\begin{barticle}
\bauthor{\bsnm{{Doschek}}, \binits{G.A.}},
\bauthor{\bsnm{{Meekins}}, \binits{J.F.}},
\bauthor{\bsnm{{Kreplin}}, \binits{R.W.}},
\bauthor{\bsnm{{Chubb}}, \binits{T.A.}},
\bauthor{\bsnm{{Friedman}}, \binits{H.}}:
\byear{1971},
\batitle{{Iron-Line Emission during Solar Flares}}.
\bjtitle{\apj}
\bvolume{170},
\bfpage{573}.
\doiurl{10.1086/151243}.
\adsurl{1971ApJ...170..573D}.
\end{barticle}
\endbibitem

\bibitem[\protect\citeauthoryear{{Freeland} and {Handy}}{1998}]{fre98}
\begin{barticle}
\bauthor{\bsnm{{Freeland}}, \binits{S.L.}},
\bauthor{\bsnm{{Handy}}, \binits{B.N.}}:
\byear{1998},
\batitle{{Data Analysis with the SolarSoft System}}.
\bjtitle{\solphys}
\bvolume{182},
\bfpage{497}.
\doiurl{10.1023/A:1005038224881}.
\adsurl{1998SoPh..182..497F}.
\end{barticle}
\endbibitem

\bibitem[\protect\citeauthoryear{{Gabriel}}{1972}]{gab72}
\begin{barticle}
\bauthor{\bsnm{{Gabriel}}, \binits{A.H.}}:
\byear{1972},
\batitle{{Dielectronic satellite spectra for highly-charged helium-like
  ionlines}}.
\bjtitle{\mnras}
\bvolume{160},
\bfpage{99}.
\adsurl{1972MNRAS.160...99G}.
\end{barticle}
\endbibitem

\bibitem[\protect\citeauthoryear{{Grineva} \textit{et~al.}}{1973}]{gri73}
\begin{barticle}
\bauthor{\bsnm{{Grineva}}, \binits{Y.I.}},
\bauthor{\bsnm{{Karev}}, \binits{V.I.}},
\bauthor{\bsnm{{Korneev}}, \binits{V.V.}},
\bauthor{\bsnm{{Krutov}}, \binits{V.V.}},
\bauthor{\bsnm{{Mandelstam}}, \binits{S.L.}},
\bauthor{\bsnm{{Vainstein}}, \binits{L.A.}},
\bauthor{\bsnm{{Vasilyev}}, \binits{B.N.}},
\bauthor{\bsnm{{Zhitnik}}, \binits{I.A.}}:
\byear{1973},
\batitle{{Solar X-Ray Spectra Observed from the `Intercosmos-4' Satellite and
  the `Vertical-2' Rocket}}.
\bjtitle{\solphys}
\bvolume{29},
\bfpage{441}.
\doiurl{10.1007/BF00150824}.
\adsurl{1973SoPh...29..441G}.
\end{barticle}
\endbibitem

\bibitem[\protect\citeauthoryear{{Henke}, {Gullikson}, and
  {Davis}}{1993}]{Hen93}
\begin{barticle}
\bauthor{\bsnm{{Henke}}, \binits{B.L.}},
\bauthor{\bsnm{{Gullikson}}, \binits{E.M.}},
\bauthor{\bsnm{{Davis}}, \binits{J.C.}}:
\byear{1993},
\batitle{{X-Ray Interactions: Photoabsorption, Scattering, Transmission, and
  Reflection at E = 50-30,000 eV, Z = 1-92}}.
\bjtitle{Atom. Data Nucl. Data Tables}
\bvolume{54},
\bfpage{181}.
\doiurl{10.1006/adnd.1993.1013}.
\adsurl{1993ADNDT..54..181H}.
\end{barticle}
\endbibitem

\bibitem[\protect\citeauthoryear{{Kelly}}{1987}]{kel87}
\begin{botherref}
\oauthor{\bsnm{{Kelly}}, \binits{R.L.}}:
1987,
{Atomic and ionic spectrum lines below 2000 Angstroms. Hydrogen through
  Krypton}.
\textit{J. Phys. Chem. Reference Data}
\textbf{16}.
\adsurl{1987JPCRD..16S....K}.
\end{botherref}
\endbibitem

\bibitem[\protect\citeauthoryear{{Lemen} \textit{et~al.}}{1984}]{lem84}
\begin{barticle}
\bauthor{\bsnm{{Lemen}}, \binits{J.R.}},
\bauthor{\bsnm{{Phillips}}, \binits{K.J.H.}},
\bauthor{\bsnm{{Cowan}}, \binits{R.D.}},
\bauthor{\bsnm{{Hata}}, \binits{J.}},
\bauthor{\bsnm{{Grant}}, \binits{I.P.}}:
\byear{1984},
\batitle{{Inner-shell transitions of Fe XXIII and Fe XXIV in the X-ray spectra
  of solar flares}}.
\bjtitle{\aap}
\bvolume{135},
\bfpage{313}.
\adsurl{1984A\%26A...135..313L}.
\end{barticle}
\endbibitem

\bibitem[\protect\citeauthoryear{{McKenzie} and {Landecker}}{1981}]{mck81}
\begin{barticle}
\bauthor{\bsnm{{McKenzie}}, \binits{D.L.}},
\bauthor{\bsnm{{Landecker}}, \binits{P.B.}}:
\byear{1981},
\batitle{{Analysis of series of solar flare X-ray spectra}}.
\bjtitle{\apj}
\bvolume{248},
\bfpage{1117}.
\doiurl{10.1086/159240}.
\adsurl{1981ApJ...248.1117M}.
\end{barticle}
\endbibitem

\bibitem[\protect\citeauthoryear{{Moser} \textit{et~al.}}{2005}]{mos05}
\begin{bchapter}
\bauthor{\bsnm{{Moser}}, \binits{M.}},
\bauthor{\bsnm{{Semprimoschnig}}, \binits{C.}},
\bauthor{\bsnm{{van Eesbeek}}, \binits{M.}},
\bauthor{\bsnm{{Pippan}}, \binits{R.}}:
\byear{2005},
\bctitle{{Comparison of results from post-flight investigations on FEP
  retrieved from the Hubble Space Telescope solar arrays and LDEF}}.
In: \bbtitle{Spacecraft Structures, Materials and Mechanical Testing 2005},
\bsertitle{ESA Special Publication}
\bseriesno{581}.
\adsurl{2005ESASP.581E..95M}.
\end{bchapter}
\endbibitem

\bibitem[\protect\citeauthoryear{{Neupert} and {Swartz}}{1970}]{neu70}
\begin{barticle}
\bauthor{\bsnm{{Neupert}}, \binits{W.M.}},
\bauthor{\bsnm{{Swartz}}, \binits{M.}}:
\byear{1970},
\batitle{{Resonance and Satellite Lines of Highly Ionized Iron in the Solar
  Spectrum Near 1.9 {\AA}}}.
\bjtitle{\apjl}
\bvolume{160},
\bfpage{L189}.
\doiurl{10.1086/180557}.
\adsurl{1970ApJ...160L.189N}.
\end{barticle}
\endbibitem

\bibitem[\protect\citeauthoryear{{Parkinson}}{1972}]{par72}
\begin{barticle}
\bauthor{\bsnm{{Parkinson}}, \binits{J.H.}}:
\byear{1972},
\batitle{{Satellite Lines in the Solar X-ray Spectrum of Helium-like
  Magnesium}}.
\bjtitle{Nature Physical Science}
\bvolume{236},
\bfpage{68}.
\doiurl{10.1038/physci236068a0}.
\adsurl{1972NPhS..236...68P}.
\end{barticle}
\endbibitem

\bibitem[\protect\citeauthoryear{{Parmar}}{1981}]{parm81}
\begin{botherref}
\oauthor{\bsnm{{Parmar}}, \binits{A.N.}}:
1981,
{Studies of X-ray Emission from Hercules X-1 (OAO1653-40) and X-ray
  Spectroscopy of Solar Flares}.
PhD thesis,
Mullard Space Science Laboratory, University College London, Holmbury St.~Mary,
  Dorking, Surrey RH5 6NT, UK.
\end{botherref}
\endbibitem

\bibitem[\protect\citeauthoryear{{Rapley}}{1976}]{rap76}
\begin{botherref}
\oauthor{\bsnm{{Rapley}}, \binits{C.G.}}:
1976,
{New techniques and observations in Soft X-ray astronomy}.
PhD thesis,
Mullard Space Science Laboratory, University College London, Holmbury St.~Mary,
  Dorking, Surrey RH5 6NT, UK.
\adsurl{1976PhDT.......170R}.
\end{botherref}
\endbibitem

\bibitem[\protect\citeauthoryear{{Rapley} \textit{et~al.}}{1977}]{rap77}
\begin{barticle}
\bauthor{\bsnm{{Rapley}}, \binits{C.G.}},
\bauthor{\bsnm{{Culhane}}, \binits{J.L.}},
\bauthor{\bsnm{{Acton}}, \binits{L.W.}},
\bauthor{\bsnm{{Catura}}, \binits{R.C.}},
\bauthor{\bsnm{{Joki}}, \binits{E.G.}},
\bauthor{\bsnm{{Bakke}}, \binits{J.C.}}:
\byear{1977},
\batitle{{Bent crystal spectrometer for solar X-ray spectroscopy}}.
\bjtitle{Rev. Sci. Instr.}
\bvolume{48},
\bfpage{1123}.
\doiurl{10.1063/1.1135211}.
\adsurl{1977RScI...48.1123R}.
\end{barticle}
\endbibitem

\bibitem[\protect\citeauthoryear{{Rice} \textit{et~al.}}{2014}]{rice14}
\begin{barticle}
\bauthor{\bsnm{{Rice}}, \binits{J.E.}},
\bauthor{\bsnm{{Reinke}}, \binits{M.L.}},
\bauthor{\bsnm{{Ashbourn}}, \binits{J.M.A.}},
\bauthor{\bsnm{{Gao}}, \binits{C.}},
\bauthor{\bsnm{{Victora}}, \binits{M.M.}},
\bauthor{\bsnm{{Chilenski}}, \binits{M.A.}},
\bauthor{\bsnm{{Delgado-Aparicio}}, \binits{L.}},
\bauthor{\bsnm{{Howard}}, \binits{N.T.}},
\bauthor{\bsnm{{Hubbard}}, \binits{A.E.}},
\bauthor{\bsnm{{Hughes}}, \binits{J.W.}},
\bauthor{\bsnm{{Irby}}, \binits{J.H.}}:
\byear{2014},
\batitle{{X-ray observations of Ca$^{19 +}$, Ca$^{18 +}$ and satellites from
  Alcator C-Mod tokamak plasmas}}.
\bjtitle{J. Phys. B Atom. Mol. Phys.}
\bvolume{47}(\bissue{7}),
\bfpage{075701}.
\doiurl{10.1088/0953-4075/47/7/075701}.
\adsurl{2014JPhB...47g5701R}.
\end{barticle}
\endbibitem

\bibitem[\protect\citeauthoryear{{Seely} and {Doschek}}{1989}]{see89}
\begin{barticle}
\bauthor{\bsnm{{Seely}}, \binits{J.F.}},
\bauthor{\bsnm{{Doschek}}, \binits{G.A.}}:
\byear{1989},
\batitle{{Measurement of wavelengths for inner-shell transitions in CA
  XVII-XIX}}.
\bjtitle{\apj}
\bvolume{338},
\bfpage{567}.
\doiurl{10.1086/167219}.
\adsurl{1989ApJ...338..567S}.
\end{barticle}
\endbibitem

\bibitem[\protect\citeauthoryear{{Seely} and {Feldman}}{1984}]{see84}
\begin{barticle}
\bauthor{\bsnm{{Seely}}, \binits{J.F.}},
\bauthor{\bsnm{{Feldman}}, \binits{U.}}:
\byear{1984},
\batitle{{Direct measurement of the increase in altitude of the soft X-ray
  emission region during a solar flare}}.
\bjtitle{\apjl}
\bvolume{280},
\bfpage{L59}.
\doiurl{10.1086/184270}.
\adsurl{1984ApJ...280L..59S}.
\end{barticle}
\endbibitem

\bibitem[\protect\citeauthoryear{{Siarkowski} \textit{et~al.}}{2016}]{sia16}
\begin{barticle}
\bauthor{\bsnm{{Siarkowski}}, \binits{M.}},
\bauthor{\bsnm{{Sylwester}}, \binits{J.}},
\bauthor{\bsnm{{B{\c a}ka{\l}a}}, \binits{J.}},
\bauthor{\bsnm{{Szaforz}}, \binits{{\. Z}.}},
\bauthor{\bsnm{{Kowali{\'n}ski}}, \binits{M.}},
\bauthor{\bsnm{{Kordylewski}}, \binits{Z.}},
\bauthor{\bsnm{{P{\l}ocieniak}}, \binits{S.}},
\bauthor{\bsnm{{Podg{\'o}rski}}, \binits{P.}},
\bauthor{\bsnm{{Sylwester}}, \binits{B.}},
\bauthor{\bsnm{{Trzebi{\'n}ski}}, \binits{W.}},
\bauthor{\bsnm{{St{\c e}{\'s}licki}}, \binits{M.}},
\bauthor{\bsnm{{H.~Phillips}}, \binits{K.J.}},
\bauthor{\bsnm{{Dudnik}}, \binits{O.V.}},
\bauthor{\bsnm{{Kurbatov}}, \binits{E.}},
\bauthor{\bsnm{{Kuznetsov}}, \binits{V.D.}},
\bauthor{\bsnm{{Kuzin}}, \binits{S.}},
\bauthor{\bsnm{{Zimovets}}, \binits{I.V.}}:
\byear{2016},
\batitle{{ChemiX: a Bragg crystal spectrometer for the Interhelioprobe
  interplanetary mission}}.
\bjtitle{Exp. Astr.}
\bvolume{41},
\bfpage{327}.
\doiurl{10.1007/s10686-016-9491-4}.
\adsurl{2016ExA....41..327S}.
\end{barticle}
\endbibitem

\bibitem[\protect\citeauthoryear{{Sylwester} \textit{et~al.}}{2005}]{jsyl05}
\begin{barticle}
\bauthor{\bsnm{{Sylwester}}, \binits{J.}},
\bauthor{\bsnm{{Gaicki}}, \binits{I.}},
\bauthor{\bsnm{{Kordylewski}}, \binits{Z.}},
\bauthor{\bsnm{{Kowali{\'n}ski}}, \binits{M.}},
\bauthor{\bsnm{{Nowak}}, \binits{S.}},
\bauthor{\bsnm{{P{\l}ocieniak}}, \binits{S.}},
\bauthor{\bsnm{{Siarkowski}}, \binits{M.}},
\bauthor{\bsnm{{Sylwester}}, \binits{B.}},
\bauthor{\bsnm{{Trzebi{\'n}ski}}, \binits{W.}},
\bauthor{\bsnm{{Baka{\l}a}}, \binits{J.}},
\bauthor{\bsnm{{Culhane}}, \binits{J.L.}},
\bauthor{\bsnm{{Whyndham}}, \binits{M.}},
\bauthor{\bsnm{{Bentley}}, \binits{R.D.}},
\bauthor{\bsnm{{Guttridge}}, \binits{P.R.}},
\bauthor{\bsnm{{Phillips}}, \binits{K.J.H.}},
\bauthor{\bsnm{{Lang}}, \binits{J.}},
\bauthor{\bsnm{{Brown}}, \binits{C.M.}},
\bauthor{\bsnm{{Doschek}}, \binits{G.A.}},
\bauthor{\bsnm{{Kuznetsov}}, \binits{V.D.}},
\bauthor{\bsnm{{Oraevsky}}, \binits{V.N.}},
\bauthor{\bsnm{{Stepanov}}, \binits{A.I.}},
\bauthor{\bsnm{{Lisin}}, \binits{D.V.}}:
\byear{2005},
\batitle{{RESIK: A Bent Crystal X-ray Spectrometer for Studies of Solar Coronal
  Plasma Composition}}.
\bjtitle{\solphys}
\bvolume{226},
\bfpage{45}.
\doiurl{10.1007/s11207-005-6392-5}.
\adsurl{2005SoPh..226...45S}.
\end{barticle}
\endbibitem

\bibitem[\protect\citeauthoryear{{Sylwester} \textit{et~al.}}{2015}]{jsyl15}
\begin{barticle}
\bauthor{\bsnm{{Sylwester}}, \binits{J.}},
\bauthor{\bsnm{{Kordylewski}}, \binits{Z.}},
\bauthor{\bsnm{{P{\l}ocieniak}}, \binits{S.}},
\bauthor{\bsnm{{Siarkowski}}, \binits{M.}},
\bauthor{\bsnm{{Kowali{\'n}ski}}, \binits{M.}},
\bauthor{\bsnm{{Nowak}}, \binits{S.}},
\bauthor{\bsnm{{Trzebi{\'n}ski}}, \binits{W.}},
\bauthor{\bsnm{{{\'S}t{\c e}{\'s}licki}}, \binits{M.}},
\bauthor{\bsnm{{Sylwester}}, \binits{B.}},
\bauthor{\bsnm{{Sta{\'n}czyk}}, \binits{E.}},
\bauthor{\bsnm{{Zawerbny}}, \binits{R.}},
\bauthor{\bsnm{{Szaforz}}, \binits{{\. Z}.}},
\bauthor{\bsnm{{Phillips}}, \binits{K.J.H.}},
\bauthor{\bsnm{{F{\'a}rn{\'{\i}}k}}, \binits{F.}},
\bauthor{\bsnm{{Stepanov}}, \binits{A.}}:
\byear{2015},
\batitle{{X-ray Flare Spectra from the DIOGENESS Spectrometer and Its Concept
  Applied to ChemiX on the Interhelioprobe Spacecraft}}.
\bjtitle{\solphys}
\bvolume{290},
\bfpage{3683}.
\doiurl{10.1007/s11207-014-0644-1}.
\adsurl{2015SoPh..290.3683S}.
\end{barticle}
\endbibitem

\bibitem[\protect\citeauthoryear{{Thompson}}{2006}]{tho06}
\begin{barticle}
\bauthor{\bsnm{{Thompson}}, \binits{W.T.}}:
\byear{2006},
\batitle{{Coordinate systems for solar image data}}.
\bjtitle{\aap}
\bvolume{449},
\bfpage{791}.
\doiurl{10.1051/0004-6361:20054262}.
\adsurl{2006A\%26A...449..791T}.
\end{barticle}
\endbibitem

\bibitem[\protect\citeauthoryear{{Waggett}}{1986}]{wag86}
\begin{botherref}
\oauthor{\bsnm{{Waggett}}, \binits{P.W.}}:
1986,
{Thematic Observations of Solar Flares}.
PhD thesis,
Mullard Space Science Laboratory, University College London, Holmbury St.~Mary,
  Dorking, Surrey RH5 6NT, UK.
\end{botherref}
\endbibitem

\bibitem[\protect\citeauthoryear{{Walker} and {Rugge}}{1971}]{wal71}
\begin{barticle}
\bauthor{\bsnm{{Walker}}, \binits{A.B.C.} \bsuffix{Jr.}},
\bauthor{\bsnm{{Rugge}}, \binits{H.R.}}:
\byear{1971},
\batitle{{Observation of Autoionizing States in the Solar Corona}}.
\bjtitle{\apj}
\bvolume{164},
\bfpage{181}.
\doiurl{10.1086/150828}.
\adsurl{1971ApJ...164..181W}.
\end{barticle}
\endbibitem

\end{thebibliography}

\end{article}

\end{document}